\documentclass[12pt,preprint]{aastex}

\usepackage{psfig}

\newcommand{\Lya}{\mbox{Ly$\alpha$}}

\newcommand{\NHEII}{\mbox{$N_\mathrm{HeII}$}}

\newcommand{\heii}{\mbox{\ion{He}{2}}} 
 
\newcommand{\instinit}{
 \setcounter{footnote}{1}
}
\newcommand{\instnew}[1]{
  \value{footnote}    
  \addtocounter{footnote}{-1}  
  \refstepcounter{footnote}
  \label{#1}
 }

\slugcomment{Accepted for publication in the \textit{The
    Astrophysical Journal}\\}

\shorttitle{The  \heii\, Gunn-Peterson Effect 
  towards \objectname[]{HE~2347-4342}}
\shortauthors{Smette, A. et al.}

\usepackage{natbib}
\begin{document}


\title{\textit{HST} STIS observations of the \heii\, Gunn-Peterson effect
  towards \objectname[]{HE~2347--4342}\altaffilmark{\ref{ack}}}


\author{Alain Smette\altaffilmark{\ref{gsfc},\ref{noao},\ref{fnrs}}}
\email{smette@astro.ulg.ac.be}
\author{Sara R. Heap \altaffilmark{\ref{gsfc}}}
\email{heap@srh.gsfc.nasa.gov}
\author{Gerard M. Williger\altaffilmark{\ref{gsfc},\ref{noao}}}
\email{williger@tejut.gsfc.nasa.gov}
\author{Todd M. Tripp\altaffilmark{\ref{princeton}}}
\email{tripp@astro.princeton.edu}
\author{Edward B. Jenkins\altaffilmark{\ref{princeton}}}
\email{ebj@astro.princeton.edu}
\and
\author{Antoinette Songaila\altaffilmark{\ref{hawaii}}}
\email{acowie@kant.ifa.hawaii.edu}

\instinit
\altaffiltext{\instnew{ack}}{Based on observations with the NASA/ESA
  Hubble Space Telescope, obtained at the Space Telescope Science
  Institute, which is operated by the Association of Universities for
  Research in Astronomy, Inc., under NASA contract  No. NAS5--26555.}
\altaffiltext{\instnew{gsfc}}{Laboratory for Astronomy and Solar Physics,
  NASA -- Goddard Space Flight Center, Code 681, Greenbelt, MD~20771, USA}
\altaffiltext{\instnew{noao}}{NOAO, P.O. Box 26732, Tucson, AZ~85716--6732}
\altaffiltext{\instnew{fnrs}}{Chercheur qualifi{\'e}, FNRS, Belgium,
  present address: Institut d'Astrophysique et de G\'{e}ophysique,
  Universit{\'e} de Li\`{e}ge, B--4000, Li\`{e}ge, Belgium}
\altaffiltext{\instnew{princeton}}{Princeton University Observatory,
  Princeton, NJ~08544, USA} 
\altaffiltext{\instnew{hawaii}}{Institute for Astronomy, 2680 Woodlawn
  Drive, University of Hawaii, Honolulu, HI~96822, USA} 
%

%
\begin{abstract}
  We     present an  \textit{HST}  STIS  spectrum   of the \ion{He}{2}
  Gunn-Peterson  effect towards \objectname[]{HE~2347-4342}.  Compared
  to  the    previous   \textit{HST}   GHRS     data    obtained    by
  \citet{Reimers97}, the STIS spectrum has a much improved resolution.
  The 2-dimensional detector also allows us to better characterize the
  sky and  dark background.  We  confirm the presence of  two spectral
  ranges  of much   reduced  opacity, the  opacity  gaps, and  provide
  improved  lower limits   on the  \ion{He}{2}  Gunn-Peterson  opacity
  $\tau_\mathrm{HeII}$ in  the high opacity  regions.  We use the STIS
  spectrum together    with  a Keck--HIRES  spectrum   covering  the
  corresponding   \ion{H}{1}   Ly-$\alpha$   forest   to  calculate  a
  1-dimensional map of  the softness $S$  of the  ionization radiation
  along the line  of sight towards \objectname[]{HE 2347-4342},  where
  $S$ is the ratio  of the  \ion{H}{1} to \ion{He}{2}  photoionization
  rates.  We find  that $S$ is generally  large but presents important
  variations, from $\sim 30$ in the opacity gaps to a 1 $\sigma$ lower
  limit   of 2300 at $z   \simeq  2.86$, in a   region  which shows an
  extremely low  \ion{H}{1} opacity over a 6.5  \AA\,  spectral range. 
  We note that a large softness parameter  naturally accounts for most
  of the large  \ion{Si}{4} to \ion{C}{4}  ratios seen in other quasar
  absorption line spectra.  We present  a simple model that reproduces
  the  shape of the   opacity gaps  in  absence  of  large  individual
  absorption lines.  We extend  the model described in  \citet{Heap00}
  to account for the presence of sources close to the line of sight of
  the background      quasar.   As an    alternative    to the delayed
  reionization model  suggested by \citet{Reimers97}, we  propose that
  the large   softness observed  at $z  \simeq  2.86$  is  due to  the
  presence of bright soft sources close to the line of sight, i.e. for
  which the  ratio between  the number  of  \ion{H}{1} to  \ion{He}{2}
  ionizing photons reaching  the  intergalactic medium is  large.   We
  discuss  these two models  and suggest  ways to discriminate between
  them.
\end{abstract}



\section{Introduction}

General considerations of the shape and intensity of the UV background
and the ionization  state  of the  intergalactic medium  indicate that
He$^{+}$ is the most abundant absorbing ion  in low-density regions of
the  Universe at $z  \sim  3$  \citep[e.g.][]{zd95,cwk+97,Miralda+00}. 
Consequently, a continuous depression  of the flux  of quasars (or any
other  sources) blueward of the  \ion{He}{2} Ly-$\alpha$ emission line
is predicted.  By similarity   with the effect expected  in \ion{H}{1}
\citep{Gunn65,Scheuer65},  this  depression is called  the \ion{He}{2}
Gunn-Peterson  effect.   One can   easily   show  that  the  bulk   of
\ion{He}{2} opacity, $\tau_\mathrm{HeII}$, must originate in clouds of
low  \ion{H}{1}    column   density,   $\log{N_\mathrm{HI}}    <   13$
\citep{cwk+97,Fardal98}; in other words, the \ion{He}{2} Gunn-Peterson
effect probes the  lowest density region of the  Universe, and most of
its  volume.  Consequently, this   effect allows  us  to probe  the UV
background ionizing  radiation.   On the other  hand, spectral regions
showing low \ion{He}{2} opacity,  or opacity gaps,  are also expected. 
They may arise either as  a consequence of additional ionizing sources
close to the line of sight  to the background quasar, in intrinsically
low density  regions,  or  in  regions dominated  by the  presence  of
shock-heated gas.

The    \ion{He}{2}  Gunn-Peterson   effect    is difficult  to   study
observationally due  to the scarcity  of lines of sight towards bright
quasars  devoid of  strong Lyman limit  systems \citep{MoJa90,PiJa93}. 
\citet{Jakobsen94}   were  the  first    to  study  the    \ion{He}{2}
Gunn-Peterson absorption using a  spectrum of  the  $z =  3.28$ quasar
\objectname[]{Q~0302-003} obtained with the Faint Object Camera on the
\textit{Hubble   Space  Telescope}         (\textit{HST}).  Using  the
\textit{Hopkins  Ultraviolet    Telescope},  \citet{Davidsen96}  found
$\tau_\mathrm{HeII}          =            1.00\pm0.07$         towards
\objectname[]{HS~1700+6416}.  \objectname[]{Q~0302-003}    was    also
observed   by  \citet{Hogan+97}   with  the  Goddard   High Resolution
Spectrograph   (GHRS),  which  revealed $\tau_\mathrm{HeII}  >   1.3$. 
Subsequently, \citet{Heap00} reported on new \textit{HST} STIS spectra
of this  quasar.  The high   quality of the  STIS spectrum  -- due, in
particular, to the 2-D nature of the FUV MAMA detector allowing a very
good background determination -- revealed regions of high opacity with
$\tau_\mathrm{HeII} \sim  4.8$ as well as  regions  of low opacity, or
opacity  gaps, extending over several  Mpc.   They also presented  the
results of a modeling of the spectra for which the opacity is based on
the redshifts,  $z$, and \ion{H}{1} column densities, $N_\mathrm{HI}$,
defined by the observed lines in the \ion{H}{1} Ly-$\alpha$ forest and
a diffuse  gas component  represented by  a randomly drawn  sample  of
lines  with $\log{N_\mathrm{HI}}   < 13$.  They  found  that  the high
opacity regions require a diffuse  gas component and  a large value of
$\eta \equiv N_\mathrm{HeII}/N_\mathrm{HI} \simeq 350$.  In turn, this
result implied   that the UV  background  has a soft spectrum,  with a
softness  parameter, $S      \equiv        \Gamma_\mathrm{HI}^{\mathrm
  J}/\Gamma_\mathrm{HeII}^\mathrm{J}    \simeq      800  $,  where the
$\Gamma_i^{\mathrm J}$ are the UV background photoionization rates for
\ion{H}{1} and \ion{He}{2},  respectively.  They also tested different
hypotheses to explain a prominent opacity gap at $z = 3.05$ and argued
that it most likely arises in a region where  helium is doubly ionized
by a discrete local source, quite possibly an AGN.
 
The present paper reports on new \textit{HST} STIS observations of the
\ion{He}{2}      Gunn-Peterson      effect     towards    the   quasar
\objectname[]{HE~2347-4342}.  Previous  \textit{HST}   GHRS spectra of
this object obtained  by \citet{Reimers97} revealed the first evidence
of a ``patchy'' intergalactic medium: their spectra showed (a) regions
of low opacity  in \ion{He}{2} with a  matching lack of  absorption in
\ion{H}{1}; (b)   regions of high opacity  in  \ion{He}{2} at the same
redshift as several  \ion{H}{1} lines; (c) regions  of high opacity in
\ion{He}{2}    at redshifts where   no  or  only   low column  density
\ion{H}{1} lines are detected. \citet{Reimers97} interpret theses
observations as evidence that re-ionization of \ion{He}{2} has not yet
taken place, being \textit{delayed} relative to \ion{H}{1}
re-ionization. 

The  new STIS observations provide  a significant improvement in terms
of signal-to-noise ratio and a   more secure background  determination
compared to the GHRS data.  The STIS spectra also have a substantially
better     spectral     resolution.    The   following   section,   \S
\ref{sec:observations},      summarizes the observations.      Section
\ref{sec:results} describes the spectrum and the opacity measurements.
In  \S  \ref{sec:fluctuations},  we      describe how   to  build    a
one-dimensional    map   of the   softness   parameter,   in   a  very
model-independent way.  In   \S \ref{sec:model}, we present   a simple
model of the opacity gaps and we expand the model described in Paper I
to take into  account several ionizing  sources close  to the line  of
sight to the quasar. We also discuss the possible  causes of the large
softness  parameter  seen in a  large  fraction of the  spectral range
covered   by   our data. In  particular,      we compare  the  delayed
re-ionization model \citep{Reimers97} to a scenario where soft sources
lead to a local increase of the  UV background softness parameter.  In
\S  \ref{sec:summary}, we provide    a summary of  the  paper,  and we
compare  our    most  important  results   with  recent  \textit{FUSE}
observations of this object \citep{KSO+01}.  Throughout this paper, we
use   the following    cosmological   parameters:    $H_\mathrm{0}   =
65~\mbox{km}~\mbox{s}^{-1}~\mbox{Mpc}^{-1}$, $\Omega_\mathrm{M} = 0.3$
and $\Lambda = 0.7$

\section{Observations}
\label{sec:observations}

In order to improve our understanding of the \ion{He}{2} Gunn-Peterson
effect shown in the \textit{HST} STIS spectra of
\objectname[]{HE~2347-4342}, we also require good spectra covering the
\ion{H}{1} Ly-$\alpha$ forest. In this section, we describe the
 \textit{HST} STIS and Keck--HIRES observations of this object.

\subsection{\textit{HST} STIS observations of the redshifted
  \ion{He}{2} region.}  
\label{sec:observations_stis}

We obtained the STIS observations of \objectname[]{HE~2347-4342} in
November 1998 during the course of two ``visits'', each five orbits
long (Program 7575).  The observations used the G140M grating which
produced a spectrum covering the wavelength interval 1145--1200~\AA .
The combined 10 exposures led to a total exposure
time of 29,000s.  The spectra have a
resolution of 0.16~\AA\, FWHM, equivalent to 3 ``lo-res'' pixels 
\citep{W+98,K+98}. The corresponding resolving power is $R = 7,300$.

The observations were reduced at the Goddard Space Flight Center with
the STIS Investigation Definition Team (IDT) version of CALSTIS
\citep{Lindler98}, which allowed us to make a customized treatment of
the background.  Such flexibility is needed in order to ensure
accurate fluxes and opacities in the Gunn-Peterson trough.

The  detector   background  in  a STIS  G140M    observation is highly
non-uniform, with a large  ``hot-spot'' of dark counts which increases
in amplitude with increasing detector temperature \citep[see e.g. Fig.
1 of ][]{Brown+00}.  The QSO  spectrum intersects this hot-spot roughly
through the middle.   Because of the  non-uniformity of the background
and the  variation of its level from  exposure to exposure, we decided
to  reduce  and combine  the spectra  following the optimal extraction
algorithm detailed by  \citet{r86}.  This method  led to a significant
improvement of  the final S/N because  it  provides optimal weights to
individual spectra,  so that  those extracted  from high  dark current
exposures contribute   proportionally  less  than    those  with   low
background.   The  only    deviation from  the   \citet{r86} algorithm
involves the estimation of the background  in each exposure, for which
we used the  following method.  Two  50--pixel wide zones were defined
above and  below the QSO  spectrum, both offset by 11  pixels from it. 
The mean  intensity of the  50 pixels of  each column of each zone was
calculated,  after  pixels   whose   intensities  were   significantly
discordant  were discarded.   This operation  produced two independent
spectra representing noisy estimates of  the sky plus dark background. 
Much smoother   versions of these  backgrounds  were then  obtained by
fitting them with a cubic spline function of 15 equally spaced nodes
using a non-linear least squares  method \citep{Bevington92}. The mean
value of the  fits to the  two backgrounds provided  us with the final
estimate of the sum of the sky and dark backgrounds.

In  \S \ref{sec:results},  we will  present very  low residual fluxes:
averaged over typically 5 \AA, they correspond to  values of the order
of                      $2.5                  ~\times                ~
10^{-17}~\mbox{ergs}~\mbox{s}^{-1}~\mbox{cm}^{-2}~\mbox{\AA}^{-1}$  at
$\sim 1170~\mbox{\AA}$.  It is thus important   to verify that  we are
indeed able to measure  such low flux values.  In particular, we  must
make  sure that the determination of  the background near the spectrum
is not affected  by  systematic effects.   Therefore, we checked   the
above routines on dark  exposures obtained over a  period of  6 months
centered on the date  of  the observations.   The temperatures  of the
detector at the  epochs these darks  were obtained had a range similar
to those of our observations of \objectname[]{HE~2347-4342}.  We found
that the routines described above consistently allowed us to determine
zero residual fluxes within the estimated measurement errors.

\begin{figure}[ht] 
\begin{center}
  \plotone{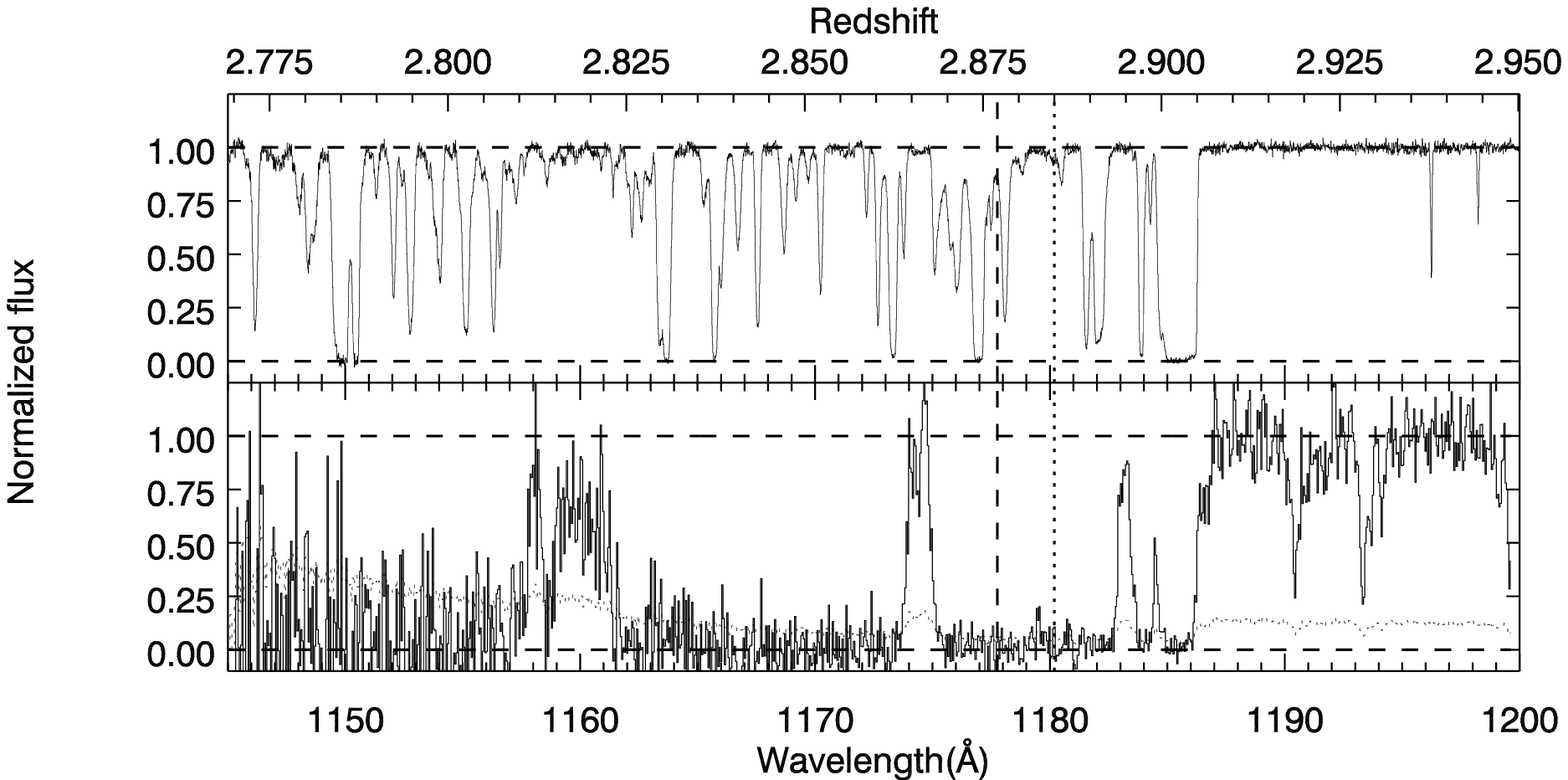} 
  \caption[]{Spectra of \objectname[]{HE~2347--4342}.
    \textit{(Top:)} Keck--HIRES  spectrum.  The  original spectrum  has
    been normalized  by a  spline  function passing   through spectral
    ranges  apparently devoid of  absorption lines.   Wavelengths have
    been multiplied by the ratio  of the wavelengths of \ion{He}{2} to
    \ion{H}{1}                 Ly-$\alpha$                       lines
    $\lambda_\mathrm{HeII}/\lambda_\mathrm{HI} = 303.7822/1215.6701$.  
    \textit{(Bottom:)} \textit{HST} STIS FUV MAMA G140M spectrum.  The
    original spectrum has been normalized assuming that the underlying
    quasar continuum has an  observed flux constant with wavelength of
    $f_\lambda                     =         2.53             ~\times~
    10^{-15}~\mbox{ergs}~\mbox{s}^{-1}~\mbox{cm}^{-2}~\mbox{\AA}^{-1}$.
    The  thin dotted line represents  the  1-$\sigma$ error array.  No
    sign of  \ion{He}{2}  Ly-$\alpha$ emission  line is   present; its
    center is  expected to fall at  $\lambda = 1180$, indicated by the
    thick vertical dotted line.  In both  panels, the vertical  dashed
    line at  $\sim  1178~\mbox{\AA}$   corresponds to $z     = 2.877$,
    redwards of which  the spectra  are likely  to be  affected by the
    known    $z_\mathrm{abs}  \simeq       z_\mathrm{em}$      systems
    \citep{Reimers97}.  Wavelengths are vacuum heliocentric.
    \label{fig:comparison_stis_hires}
    }
\end{center}
\end{figure}

The other steps of the reduction process exactly follow the algorithms
described in \citet{Lindler98}. 
The bottom panel of Fig. \ref{fig:comparison_stis_hires} shows
the final  spectrum, normalized to the continuum flux of the QSO,
assumed to be a constant whose value is determined
in \S \ref{sec:continuum}.

\subsection{Keck--HIRES  observations of 
  the \ion{H}{1} Ly-$\alpha$ forest.}
\label{sec:observations_keck}

The Keck--HIRES spectrum was obtained in one Keck {\sc I} Telescope run
in 1997, October 1-3, and is the sum of 10,  40-minute exposures, 5 in
each of two  echelle  settings to  give complete  wavelength coverage,
which leads   to considerable overlap in  the   center of  the echelle
format. The $1\farcs14 \times 7\arcsec$ slit was employed resulting in
a resolving power $R=37,000$.

The metal-line systems in this spectrum were included in a
recent analysis of high $z$ metal-line ratios in the Ly-$\alpha$ forest
\citep{Songaila98}, which also contains a general description of the
data taking and reduction procedures.  The wavelengths in the final
spectrum are   vacuum heliocentric. The normalized spectrum is presented
in the top panel of Fig. \ref{fig:comparison_stis_hires}.

For each line $i$ in the HIRES spectrum, the \ion{H}{1} column density
$N_\mathrm{HI}^i$ and  Doppler  parameter  $b^i$ were  evaluated using
FITLYMAN \citep{fb95}.  These  lines  constitute the   `observed' line
list,  which  will  be   used in    our  modeling of   the \ion{He}{2}
Gunn-Peterson trough, described in \S \ref{sec:model}.

\section{Observational Results}
\label{sec:results}

In this section,  we  describe the STIS spectrum   of \objectname[]{HE
  2347-4342}  shown in    Fig.   \ref{fig:comparison_stis_hires}.  For
purposes of description,    this spectrum can  be   divided into three
regions:   \textit{(1)}    the    region  redward   of    $\lambda   =
1187~\mbox{\AA}$, which we use to  determine the continuum flux of the
QSO; \textit{(2)}   the region covered   by  a complex $z_\mathrm{abs}
\simeq  z_\mathrm{em}$ absorption system \citep[][]{WCS81,FWP+86,MJ87}
at the   location  where the   spectrum   should be  affected by   the
`proximity   effect';  \textit{(3)}   the  region   affected  by   the
\ion{He}{2}  Gunn-Peterson   effect.    We    now examine  the    main
characteristics of each of these regions.

\subsection{The continuum redward of $\lambda = 1187~\mbox{\AA}$.}
\label{sec:continuum}

For a given object, measuring  the opacity in the Gunn-Peterson trough
mainly depends on  the throughput of the telescope  and the absence of
systematic effects at low fluxes, which we demonstrate in the previous
section. It also requires a good estimate of the underlying continuum.
In the absence of  published data blueward  of the STIS  spectrum, our
only  means to  estimate the  continuum   level is to  extrapolate the
behavior of the spectral region redward of $\lambda = 1187~\mbox{\AA}$
towards the blue.   However, we must  also  make sure that a  possible
\ion{He}{2} emission  line does not lead  to  a local increase  in the
flux in that spectral range.   \citet{Reimers97} measured the redshift
of \objectname[]{HE~2347-4342}  to be  $  z = 2.885\pm0.005$  from the
observed   wavelength of the  \ion{O}{1} $\lambda1304$  emission line. 
Since the systemic  velocity of  the \ion{He}{2} Ly-$\alpha$  emission
line has  never been measured  to our  knowledge, we  can estimate the
expected range of its observed wavelength from the systemic velocities
of other emission lines in the Keck--HIRES  spectrum.  We find that the
velocities  of the  emission  line peaks  relative  to their  expected
location  for  $ z = 2.885$  are  bracketed by the \ion{C}{4} emission
line, which is  blueshifted by $\simeq 1,670~\mbox{km}~\mbox{s}^{-1}$,
and the   \ion{N}{5} emission  line,  which  is redshifted  by $\simeq
2,900~\mbox{km}~\mbox{s}^{-1}$ -- however, we note that the \ion{N}{5}
emission line may be affected by a  broad absorption line, so that its
peak may appear at a  wavelength longer than  its intrinsic location.  
The peak of the \ion{He}{2} emission line is therefore expected to lie
at $\lambda  = 1180\pm3~\mbox{\AA}$,  well  within the  spectral range
covered   by  the  $z_\mathrm{abs}   \simeq z_\mathrm{em}$  absorption
system. Given this and the fact  that the \ion{He}{2} emission line is
weak in  the  spectra of  \objectname[]{Q~0302-003} \citep{Heap00} and
\objectname[]{PKS~1935-692} \citep{Anderson99},  we are confident that
the     measurement  of   the  continuum    redwards   of  $\lambda  =
1187~\mbox{\AA}$ is little affected by the \ion{He}{2} emission line.

Redward   of  1187 \AA, the   spectrum  presents a  continuum spectrum
interrupted    most   notably  by  a  pair     of Galactic \ion{Si}{2}
$\lambda1190,1193$  doublets  (at    heliocentric   velocities   $v  =
+7.5\pm2.5$ and $60.7\pm2.5~\mathrm{km}~\mathrm{s}^{-1}$), as well  as
some   weak   absorption  lines,  most      probably  Ly-$\beta$ lines
corresponding to low-$z$ Ly-$\alpha$ lines. Away from these absorption
lines, the  spectrum appears very flat (in  $f_\lambda$), with  a mean
flux                   of           $2.53\pm0.03              ~\times~
10^{-15}~\mbox{ergs}~\mbox{s}^{-1}~\mbox{cm}^{-2}~\mbox{\AA}^{-1}$,
measured in the spectral regions defined by $1187.1 < \lambda < 1189.9
\mbox{\AA}$ and $1194.6  < \lambda <  1198.7 \mbox{\AA}$.   Comparison
with the GHRS  spectrum  obtained by \citet{Reimers97} indicates  that
the quasar flux was $\sim$ 25\% lower  during our observations than it
was   in June  1996.    For completeness, we   note that  the Galactic
reddening    towards   \objectname[]{HE  2347-4342}    is   very  low,
$E(\mathrm{B}-\mathrm{V}) =  0.014$ \citep{sch98}. (This value is less
than  half     the      one  used    by    \citet{Reimers97},   i.e.   
$E(\mathrm{B}-\mathrm{V}) =  0.0387$, which   was inferred  from   the
\ion{H}{1} column density  measurement by \citet{Stark+92}.)  Assuming
$A_\mathrm{V}/E(\mathrm{B}-\mathrm{V})  = 3.08$  and a mean  Milky Way
extinction law  given  by the analytical expression  of \citet{pei92},
the mean flux  corrected  from Galactic extinction in   the wavelength
range    $1186   -    1200$~\AA\, is   therefore    $2.57\pm0.03 \times
~10^{-15}~\mbox{ergs}~\mbox{s}^{-1}~\mbox{cm}^{-2}~\mbox{\AA}^{-1}$.   
Furthermore,  the extinction varies only by  $\sim 1\%$ over the whole
range covered by the STIS spectrum.

Examination of the FOS and GHRS spectra presented by \citet{Reimers97}
indicates  that the \ion{He}{2}  Gunn-Peterson trough is just blueward
of  the recovery zone of  the continuous absorption  by  2 Lyman limit
systems at a mean redshift $z  = 2.735$.  More precisely, they measure
the optical depth at the \ion{H}{1} Lyman limit to be $\tau \sim 1.6$,
which  implies that the continuum absorption  changes by less than 1\%
in the wavelength range covered  by our STIS spectrum.  Unfortunately,
despite this,  existing data do not allow  a reliable determination of
the underlying, `true', unabsorbed quasar continuum slope due to their
relatively low signal-to-noise ratio.

Therefore,   we  performed a  linear  fit  to  our  STIS  data  in the
wavelength ranges $1187.1 < \lambda < 1189.9 \mbox{\AA}$ and $1194.6 <
\lambda <  1198.7 \mbox{\AA}$, apparently devoid  of absorption lines. 
We find that a flat continuum (in  $f_\lambda$) is compatible with our
data.  In the following, we have therefore  assumed that the continuum
is indeed flat.  The corresponding normalized spectrum is presented in
the bottom panel  of Fig.  \ref{fig:comparison_stis_hires}.   However,
the  error on   the slope is  such   that extrapolating the  result to
1145\AA\, at the  bluest range of our  spectrum indicates  that it may
underestimate the true  value by at most 10\%  at  that wavelength and
proportionally smaller at  longer  wavelengths.  We obtain  a  similar
result  using the lower    signal-to-noise GHRS spectrum fitted   over
additional  spectral ranges  redwards of  1200~\AA.  Consequently, the
mean normalized flux  measured  below (\S   \ref{sec:high_opacity} and
Table \ref{tab:opacity}) could be over-estimated by at most $10$\%, so
that the  opacities in the large  opacity regions, which we are mainly
interested in, could be under-estimated by at  most 0.1. Therefore, as
stated earlier, our ability to measure high  opacity depends mainly on
the low  systematic errors in   the background level made possible  by
STIS, whereas the continuum definition is less critical.

\subsection{The  absence of the \ion{He}{2} proximity effect.}
\label{sec:z_abs_z_em}

Observations of the \ion{H}{1}    Ly-$\alpha$ forest reveal  that  the
number and column densities of  \ion{H}{1} lines decrease close to the
redshift of the background quasar. This effect, known as the proximity
effect, is most probably due to an increase  in the ionizing radiation
intensity close to   the quasar \citep[cf.][for   a review]{RAU98}.  A
similar effect is expected in \ion{He}{2}, which should be easily seen
in the spectra of bright QSOs. Indeed, from Eq.  4 in \citet{zd95}, we
would expect  that the proximity profile would  extend over  more than
$\sim 30~\mbox{\AA}$ -- a large portion of  the STIS spectrum -- since
the   luminosity    of  \objectname[]{HE~2347-4342}   is    $L  \simeq
1.4\times10^{31}~\mbox{ergs}~\mbox{s}^{-1}~\mbox{Hz}^{-1}$   at    the
\ion{He}{2} Lyman  limit.  However, \citet{Reimers97}  found that  the
spectrum shows no evidence for a \ion{He}{2} proximity effect.
Instead, the GHRS and the STIS spectra reveal  regions of high opacity
near the redshift of the  quasar.  At the corresponding wavelengths in
the  Keck spectrum, a   series of interesting \ion{H}{1},  \ion{C}{4},
\ion{N}{5}  and  \ion{O}{6} absorption  lines    can be  seen.    More
surprisingly, the  \ion{N}{5} and especially  \ion{C}{4} lines present
evidence of  ``line--locking''' (Smette \&  Songaila, in preparation),
i.e., the velocity separation  between two different components of the
complex   appear equal  to  the  velocity separation  between the  two
components   of  the  \ion{C}{4}  doublet \citep[e.g.][and  references
therein]{WCS81}.  This effect results from  radiation pressure and can
only be produced by material  located physically  close to the  region
responsible for the  QSO continuum.   The GHRS  and the STIS   spectra
therefore   suggest that absorption   from this $z_\mathrm{abs} \simeq
z_\mathrm{em}$ system is also present in \ion{He}{2}.

\citet{Reimers97} suggested that  the absence of  the proximity effect
in \ion{He}{2} is actually due to the presence of this $z_\mathrm{abs}
\simeq z_\mathrm{em}$  system.  The  large \ion{He}{2} column  density
$N_\mathrm{HeII}$ required  to  explain the  observed  opacity in  the
$z_\mathrm{abs} \simeq z_\mathrm{em}$  complex in  the GHRS and   STIS
spectra also implies  a    large   \ion{He}{2} continuum opacity.      
Consequently, the  number of \ion{He}{2}  ionizing photons effectively
escaping from the background  quasar is reduced by  a large amount.  A
conservative way to estimate the continuum opacity can be made. First,
we assume  that the gas  has a `box-car'  velocity dispersion. In this
case, a lower limit to the \ion{He}{2} column density distribution can
be obtained by  measuring the  effective opacity $\tau_\mathrm{eff}  =
-\ln{\bar{I}}$  in  a    trough     of the   $z_\mathrm{abs}    \simeq
z_\mathrm{em}$ system.  Here, $\bar{I}$ is the mean normalized flux in
the spectral range of interest.  We find that  in the range $1184.80 <
\lambda        <      1186.10~\mbox{\AA}$,    $\tau_\mathrm{eff}     =
4.5^{+2.3}_{-0.7}$.   Therefore, we can  infer that $N_\mathrm{HeII} >
2.4 \times  10^{18}~\mbox{cm}^{-2}$.  As a  lower limit to the opacity
at    the   \ion{He}{2}   Lyman-limit    is   actually     given    by
$\tau_\mathrm{eff}$ itself, the   transmitted flux at  the \ion{He}{2}
Lyman-limit  escaping from  the   clouds causing   the $z_\mathrm{abs}
\simeq z_\mathrm{em}$  system   is  certainly less  than   2\%  of the
continuum flux.  This value  is sufficient to reduce  the size  of the
proximity profile to  $\sim 5~\mbox{\AA}$.  Since the  $z_\mathrm{abs}
\simeq   z_\mathrm{em}$ system is  better represented  by a complex of
lines, we can reasonably assume that the \ion{He}{2} column density is
larger  than the value given above,  reducing further the  size of the
proximity  profile. Therefore, we   can  confirm that the presence  of
$z_\mathrm{abs}  \simeq   z_\mathrm{em}$ complex in    the spectrum of
\objectname[]{HE~2347-4342} is   likely   to cause the   absence  of a
proximity effect.  In  \S \ref{sec:summary}, we  discuss the potential
impact of such systems on the general UV background.

\subsection{High and low opacities 
  in the \ion{He}{2} Gunn-Peterson spectrum}
\label{sec:high_opacity}

The presence of  the $z_\mathrm{abs} \simeq  z_\mathrm{em}$ absorption
systems makes it difficult  to determine exactly the  highest redshift
for which the intergalactic background radiation is the only influence
of the  \ion{He}{2}  ionization  fraction.  The two   lowest  redshift
\ion{C}{4} doublets that appear to be mutually line-locked are at $z =
2.87802$ and $ 2.88450$.  We thus suggest that the maximal redshift to
be considered for the study of the \ion{He}{2} Gunn-Peterson effect is
$z = 2.877$  (corresponding  to  $\lambda = 1177.76~\mbox{\AA}$    for
\ion{He}{2} Ly-$\alpha$), slightly  lower than the  value $z =  2.885$
chosen by \citet{Reimers97}.

\begin{figure}[ht]
  \begin{center}
    \plotone{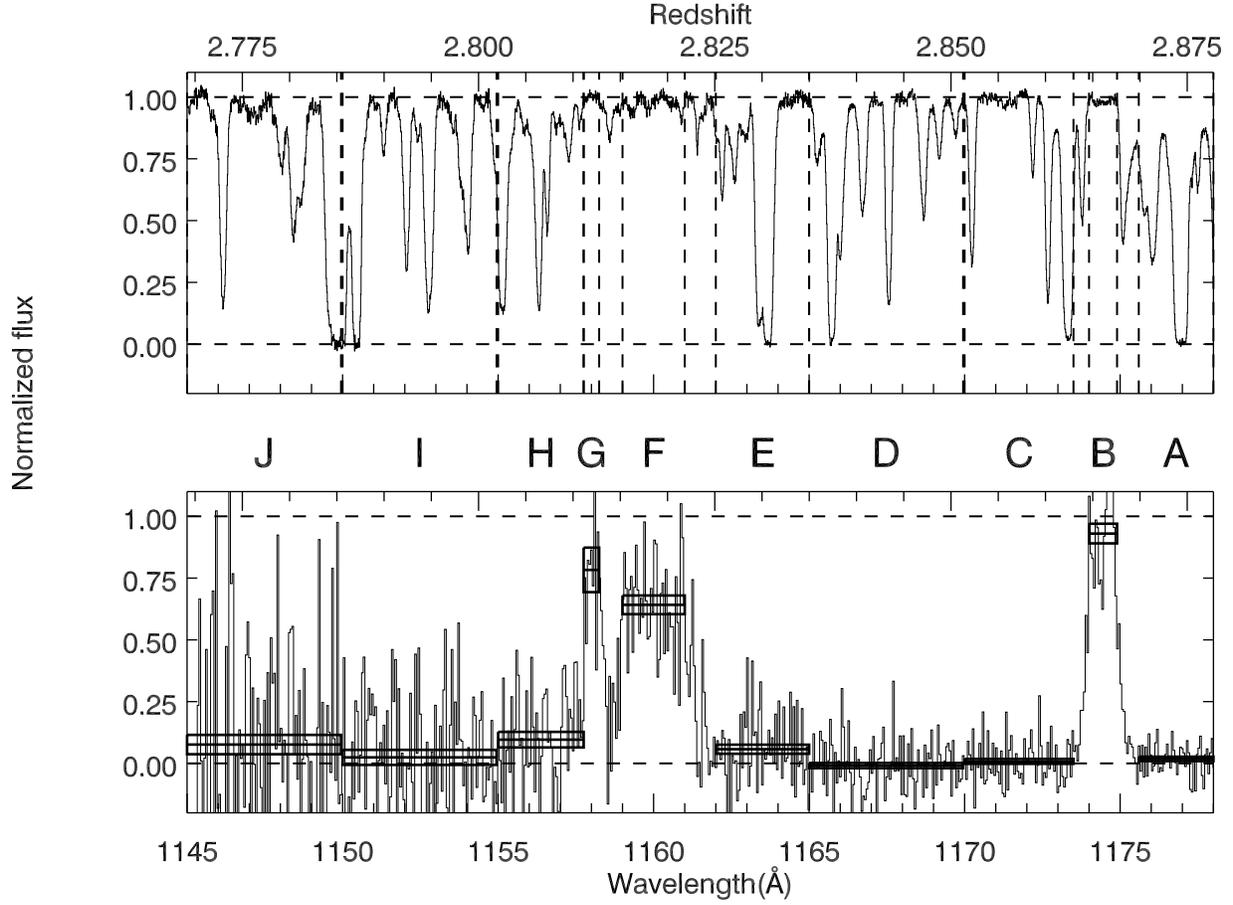} 
    \caption{Normalized STIS spectrum of HE~2347--4342. We have also
      represented by a rectangle   the  mean and uncertainty   of  the
      transmitted flux for each  of   the spectral ranges defined   in
      Table~\ref{tab:opacity}.  For   each    rectangle, the     width
      represents the wavelength range of the corresponding region, the
      central    line represents its   mean   transmitted flux and its
      thickness is equal to twice the measurement uncertainty.  }
    \label{fig:normalized_spectrum}
  \end{center}
\end{figure}

Following  \citet{Heap00}, we  measure    the mean   normalized   flux
$\bar{I}$ and its standard deviation  $\sigma_I$ in different spectral
ranges of the spectrum.  Table \ref{tab:opacity} lists the wavelengths
used to  define  these ranges as well  as  the corresponding values of
$\bar{I}$ and $\sigma_I$.  These values  are also graphically shown in
Fig.  \ref{fig:normalized_spectrum}.  Table   \ref{tab:opacity}   also
lists the   inferred  Gunn-Peterson  opacities   $\tau_\mathrm{HeII} =
-\ln{\bar{I}}$, and their   corresponding  1  sigma  upper  and  lower
limits.    If $\bar{I}$  is   negative,  the reported  lower  limit to
$\tau_\mathrm{HeII}$   is    $-\ln{\sigma_I}$.       As discovered  by
\citet{Reimers97},  the \ion{He}{2} Gunn-Peterson spectrum has regions
of high and low opacities.

At this  stage, it is  interesting to note that  the region $S$ in the
spectrum of  \objectname[]{Q~0302-003} \citep{Heap00} covers the range
$2.77   <     z  <  2.87$,   as     the  whole   STIS   spectrum    of
\objectname[]{HE~2347-4342}.    Although     the        spectrum    of
\objectname[]{Q~0302-003}   is  noisy  in  that region,   it   shows a
relatively constant flux and no opacity gap. A value $\tau_\mathrm{HeII}
= 1.88$ was inferred, which is much smaller  than the opacity measured
at higher redshift  towards the same  object and also the opacity
measured in most regions ($A$, $C$, $D$, $I$,  $J$) in the spectrum of
\objectname[]{HE~2347-4342}. \citet{Heap00}  noted this apparent sharp
decrease in \ion{He}{2} Gunn-Peterson opacity below $z  = 3$ along the
line of sight   towards \objectname[]{Q~0302-003} but suggested  that,
indeed, it  may  not be representative  of other  lines   of sight.  A
possible cause for the low  opacity of the region  $S$ in the spectrum
of \objectname[]{Q~0302-003} in presented in \S \ref{sec:model}.

Spectral range $C$ in  the spectrum of  \objectname[]{HE~2347-4342} is
of special interest because of its high opacity, $\tau_\mathrm{HeII} =
4.80_{-0.80}^{+\infty}$, even  though the  region in the corresponding
\ion{H}{1} spectrum  shows a large  spectral range  with no absorption
lines. We will show in \S  \ref{sec:model} that this spectral range is
even more peculiar than the comparable Dobrzycki-Bechtold void seen in
\objectname[]{Q~0302-003}, for which a large opacity in \ion{He}{2} is
measured despite a dearth of large column density \ion{H}{1} lines.

The opacity  gaps   at $z =  2.817$  and  2.866 \citep[called  `voids'
by][]{Reimers97} extend  over 13 and  7 comoving  Mpc and the  minimal
values of their  opacity are 0.44  and 0.07, respectively.  The latter
opacity gap   seems  associated with  an  absorption  system at   $z =
2.86237$ located  at   a velocity $v    = 280~\mbox{km}~\mbox{s}^{-1}$
blueward of  it.  It shows \ion{C}{4} at  $z  = 2.86246$ and $2.86272$
with $\log{N_\mathrm{CIV}} =  12.8$ and $12.1$, respectively. It  also
shows \ion{O}{6} with a -- relatively large -- total column density of
$\log{N_\mathrm{OVI}} = 13.6$ and a narrow Doppler $b$ parameter of 16
~  $\mbox{km}~\mbox{s}^{-1}$,  so  that the  logarithm  of the  column
density ratio $\log{N_\mathrm{CIV}/N_\mathrm{OVI}} = -0.6$ indicates a
large ionization  parameter.  The total  $\log{N_\mathrm{HI}} = 14.3$. 
The  suggestion   by  \citet{Heap00}    that \ion{C}{4}   systems  are
associated with opacity  gaps is therefore strengthened.  However,  we
note that  no absorption    system   is present within    $\sim  1000~
\mbox{km}~\mbox{s}^{-1}$ of the opacity gap at the lower redshift $z =
2.817$.

\begin{figure}[ht] 
\begin{center}
  \plotone{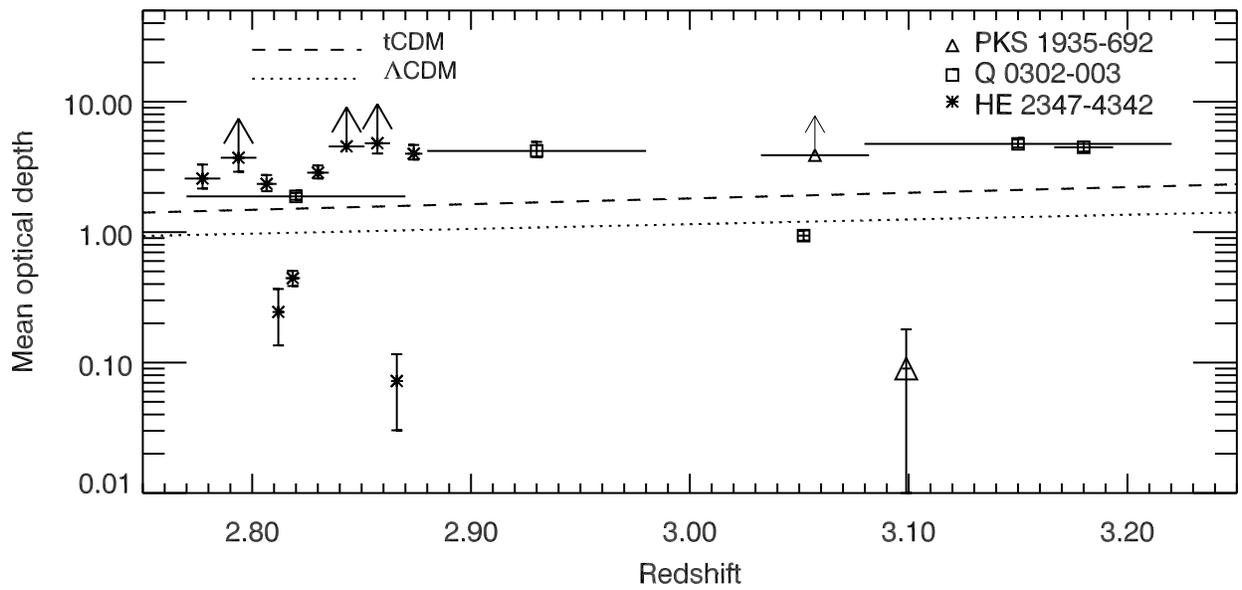} 
  \caption[]{Measurements of the \ion{He}{2}
    Gunn-Peterson opacity at $z > 2.75$.  The value measured by
    \citet{Davidsen96} towards \objectname[]{HS~1700+6416} is
    $\tau_\mathrm{HeII} = 1.0\pm0.07$ over the range $2.2 < z < 2.6$.
    For comparison, the dashed and dotted lines show the optical
    depths predicted by 
    two   models  considered by \citet{mbm+00} that bracket
    the range of models in their study.
    \label{fig:opacity_z}
    }
\end{center}
\end{figure}

In Figure \ref{fig:opacity_z}, we   compare the opacities  measured at
different  redshifts  in  the  spectra of  \objectname[]{HS~1700+6416}
\citep{Davidsen96},     \objectname[]{Q~0302-003}      \citep{Heap00},
\objectname[]{PKS~1935-692} \citep[][  the reported  values come from
our  own, optimal reduction of  the whole  data  set, cf.  Williger et
al.,  in preparation]{Anderson99} and \objectname[]{HE~2347-4342}
with   the results of hydrodynamical   simulations  based on different
cosmological models.  The two curves bracket the \ion{He}{2} opacities
of all the models calculated by \citet{mbm+00}.   It is clear that the
measured opacities are much larger  than predicted.  As concluded also
by \citet{Heap00}, we  will argue later  that these simulations  use a
model of the UV background radiation that is much too hard.

\section{A one-dimensional map of the softness parameter}
\label{sec:fluctuations}

\subsection{A method to determine the softness parameter at a given
  redshift}
\label{sec:method}

In this section, we present a method that aims at providing the value
of the softness parameter $S$ of the ionizing radiation, defined as
the ratio of the \ion{H}{1} to \ion{He}{2} photoionization rates,
\begin{equation}
S \equiv \frac{\Gamma_\mathrm{HI}}{\Gamma_\mathrm{HeII}},
\label{eq:softness_parameter}
\end{equation}
as a  function of redshift.  We  prefer this definition to another one
given by the ratios of the UV background intensities at the \ion{H}{1}
and      \ion{He}{2}    Lyman     limits,    $S_\mathrm{L}      \equiv
J_\mathrm{HI}/J_\mathrm{HeII}$,  as  Eq.   \ref{eq:softness_parameter}
accounts for the shape of  the UV background spectrum.  Therefore,  we
can actually  build a one-dimensional map of  the  fluctuations of the
softness parameter as  a function of redshift along  the line of sight
to  the    background       quasar   studied,   in       this    case,
\objectname[]{HE~2347-4342}.  In practice,  the resulting map provides
the softness parameter for the different spectral ranges defined above
(\S  \ref{sec:high_opacity}), which correspond  to a redshift interval
comparable  to  the  mean    redshift  separation between     any  two
$\log{N_\mathrm{HI}} = 14$  Lyman  forest lines.   Such a  map  can be
compared with the values from other observations, as, for example, the
\ion{Si}{4}       to      \ion{C}{4}       ratio       (cf.         \S
\ref{sec:fluctuations_siiv_civ}). We emphasize that such a map is very
model-independent, in  the  sense that we make   no hypothesis  on the
origin of the softness  parameter at a  given redshift. We  leave such
interpretation to \S \ref{sec:model}.

\subsubsection{Column density ratio $\eta$}

The  following method,  as well  as the models   described later in \S
\ref{sec:model}, assumes  that  \Lya\, clouds at  $z  \sim  3$  are in
photoionization  equilibrium, with hydrogen  mostly ionized and helium
mostly doubly ionized  \citep[e.g.][ and references  therein]{Heap00}. 
We also assume that  the ratio of column  densities of \ion{He}{2} and
\ion{H}{1} is given by
\begin{equation}
  \label{eq:n_heii__n_hi}
  \eta \equiv
  \frac{N_\mathrm{HeII}}{N_\mathrm{HI}} = 
  \frac{n_\mathrm{He}}{n_\mathrm{H}}~
  \frac{\alpha_\mathrm{HeII}}{\alpha_\mathrm{HI}}~
  \frac{\Gamma_\mathrm{HI}}{\Gamma_\mathrm{HeII}},
\end{equation}
where 
$\alpha_i$ is the recombination coefficient of ion $i$ (\ion{H}{1} or
\ion{He}{2}), 
$\Gamma_i$ is its ionization rate, and 
$n_\mathrm{He}$, $n_\mathrm{H}$ are the total number densities of
helium and hydrogen respectively. As in \citet{Heap00}, we also assume 
full turbulent broadening, so that the ratio of the \ion{He}{2} and
\ion{H}{1} lines Doppler parameters, $\xi \equiv b_\mathrm{HeII}/
b_\mathrm{HI} = 1$. Lower values,  as $\xi = 0.8$ used by
\citet{SHC95}, lead to  values of the softness parameter even larger
than the ones we derive below.
Note that for clouds with high \NHEII, self-shielding and emissivity
of the clouds start to be substantial, and this simple equation is no
longer valid.
For $n_\mathrm{He}/n_\mathrm{H} = 0.082$ as expected from Big-Bang
nucleosynthesis \citep[e.g.][]{Tytler+00} and
$\alpha_\mathrm{HeII}/\alpha_\mathrm{HI} =5.418 $
\citep[e.g.][]{OST89}, 
the equation reduces to 
\begin{equation}
\label{eq:eta_44}
\eta = 
0.44 \, \frac{\Gamma_\mathrm{ HI}}{\Gamma_\mathrm{HeII}},~~~ \mbox{or}  
\end{equation}
\begin{equation}
  \label{eq:eta_s}
  \eta = 0.44 ~ S.
\end{equation}

The photoionization rates are given by:
\begin{equation}
  \label{eq:ionization_rate}
  \Gamma_i = \int_{\nu_i}^\infty~
  \frac{f_\nu}{h \nu}~\sigma_\nu~{\rm d}\nu,
\end{equation}
where $f_\nu$  is  the total  (UV  background and  all  other sources)
ionizing  photon flux at  frequency $\nu$, $\sigma_\nu \simeq \sigma_i
~(\nu/\nu_i)^{-3}$       is   the   photoionization      cross-section
\citep[e.g.][]{OST89}, and $\nu_i$ is the frequency of the Lyman limit
of ion $i$.\\

It immediately follows  from Eq.  \ref{eq:eta_s} that  determining the
best value  for $\eta$ to match  the observations  also constrains the
softness parameter $S$.  However, any spectral range of high opacity
in the Gunn-Peterson trough only allows us  to provide a lower limit to
$\eta$, which in turn leads to a lower limit to $S$.

In order to determine $\eta$, we must assess the amount of \ion{H}{1}.
As in \citet{Heap00}, we use a `combined' line list, which is a merger
of the observed line list (cf.  \S \ref{sec:observations_keck}) and a
`simulated' line list consisting of lines with $\log{N_\mathrm{HI}} <
13$ obtained by simulations \citep[cf.][for details]{Heap00}.  Then,
for each value of $S$ considered, we can built a corresponding
\ion{He}{2} combined line list. The corresponding Voigt profiles are
then calculated.  The resulting spectrum is obtained after
convolution with the \textit{HST}  STIS instrumental line spread
function  \citep{l00}. For 
each of the spectral ranges defined above and for each simulated
spectrum (characterized by the value of $S$), we can estimate the mean
normalized flux $\bar{I}(S)$ and the corresponding opacity 
$\tau_\mathrm{HeII}(S)$.

Note however that the derived values for the softness parameter are
likely to be underestimated. 
Indeed, even if only weak or no corresponding \ion{H}{1} lines can be
seen, a large \ion{He}{2} opacity usually suggests that the gas
density is not especially small compared to other regions.
Instead, it probably indicates that the
\ion{H}{1} photoionization rate is larger in that region. Therefore,
the neutral fraction in \ion{H}{1} is probably smaller for the clouds
observed in that region compared to the mean Ly-$\alpha$ forest at the
same redshift.  Since our `simulated' line list is based on the mean
characteristics of the Ly-$\alpha$ forest at the given redshift, they
constitute a set of lines caused by clouds with a mean value of the
neutral fraction. Consequently this method is not self-consistent.  A
better analysis would require that the neutral fraction for the clouds
in the `simulated' line list be consistent with the neutral fraction
of the clouds in the `observed' line list. Such an approach is out of
the scope of this paper. However, we can infer what its effect would
be.  
The  mean number density  of lines at   a given column  density in the
`simulated' line list is expected to be smaller  in a region where the
large    \ion{H}{1} photoionization rate is   larger  than in the mean
Ly-$\alpha$ forest, as it is the case in the `observed' line list.
Therefore, the mean \ion{He}{2} opacity
produced by these lines would be reduced. Consequently, in order to
account for the observed \ion{He}{2} opacity, the softness parameter
would have to be larger.

\subsection{Evidence for a soft and fluctuating  UV  background}
\label{sec:fluctuations_siiv_civ}

\begin{figure}[ht]
  \begin{center}
  \plotone{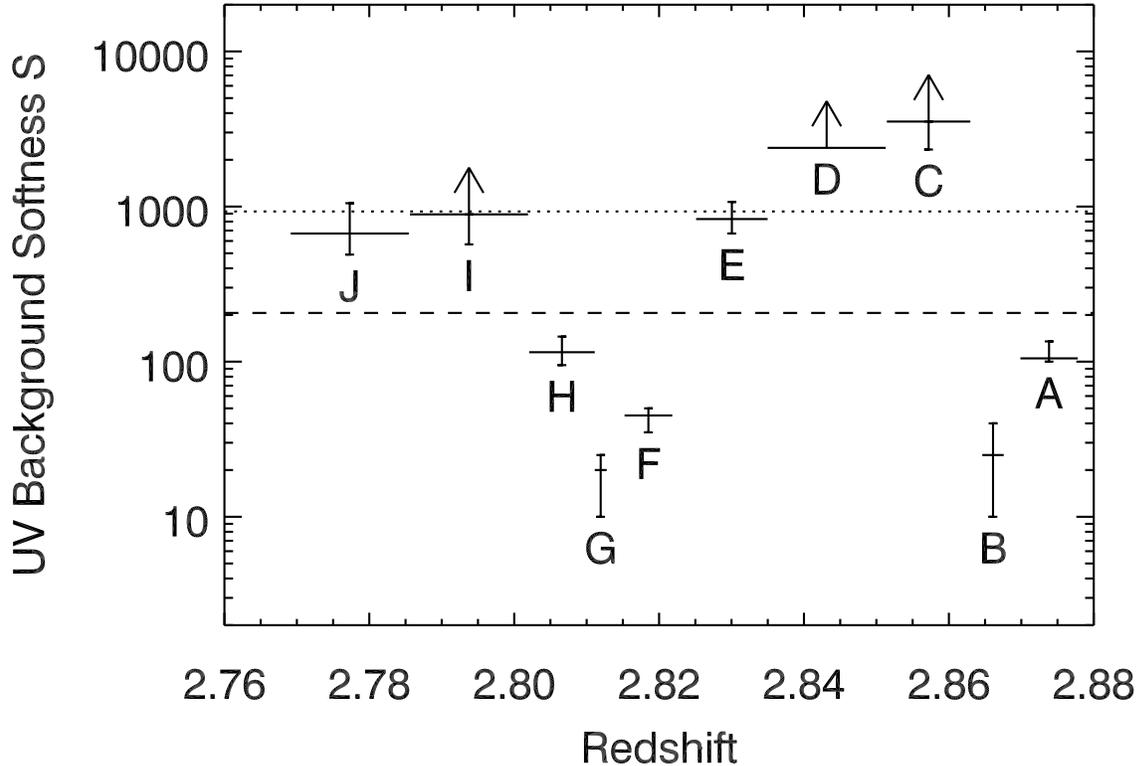} 
    \caption[]{Softness parameter for different spectral ranges in the
      spectrum of \objectname[]{HE~2347--4342}.   The  width  of  each
      symbol  covers the redshift range   of the corresponding region. 
      Arrows represent 3  $\sigma$ lower limits, i.e., they correspond
      to spectral  ranges  over which the  mean   normalized flux $\bar{I} <
      3~\sigma_\mathrm{I}$.  The  dashed line  represents the softness
      parameter of  the  model calculated  by  \citet{Madau99} at $z =
      2.76$.  The  dotted line gives the  softness parameter of one of
      the models calculated by \citet{Fardal98}, for which the stellar
      contribution is  fixed at 912 \AA\,  to have an emissivity twice
      that of the quasars (cf. text).
        \label{fig:softness_vs_z}
        }
  \end{center}
\end{figure}

Figure \ref{fig:softness_vs_z}  shows the  resulting 1-dimensional map
of   the   softness  parameter   along the   line    of sight  towards
\objectname[]{HE~2347--4342}.  The error  bars parallel to the  x-axis
indicate the redshift  ranges of the   corresponding regions. For  the
spectral ranges over which the mean normalized  flux $\bar{I} < 0$, we
used an  upward    arrow  representing   a  1   $\sigma$  lower  limit
$\sigma(\bar{I})$; its  bottom line represents  the softness parameter
corresponding to a mean normalized flux  equal to $1 \sigma(\bar{I})$. 

The only spectral ranges where a  hard ionizing radiation is required,
with $S$  lower   than  $\simeq 200$,   are the   two opacity gaps.    
Otherwise, it is clearly apparent  from Fig.   \ref{fig:softness_vs_z}
that    the softness parameter is   usually   much larger   than 200.  
\citet{Fardal98} (see their Fig.   6) present other  models of  the UV
background spectrum.   In  particular, we  note their  source model Q1
with  spectral  index  $s  =  1.8$,  a  stellar   contribution,  their
absorption  model A2 and  an allowance   for  cloud re-emission.   The
stellar contribution is fixed at 912 \AA\, to have an emissivity twice
that  of  the quasars  and is  considerably softer than   the model of
\citet{Madau99}  or other   models   without  a significant    stellar
contribution.  This model of Fardal et al.  has $S= 930$ and was found
to be consistent   with the observed   opacity on  the  line  of sight
towards \objectname[]{Q~0302-003}   \citep{Heap00},    even   in   the
Dobrzycki-Bechtold void region which presents a  low number density of
Ly-$\alpha$  lines. However, Figure \ref{fig:softness_vs_z} also shows
that even such a  soft background is too hard  for the spectral ranges
$C$ and  $D$    at  $z  \simeq   2.845$  and   2.855, respectively.    
Alternatively, the large softness  parameters could be due to  delayed
\ion{He}{2} re-ionization, as advocated by \citet{Reimers97}.  In this
case,  the corresponding regions  have  yet to  be  suject to a  large
amount  of  \ion{He}{2}  ionizing  photons,   so that  the \ion{He}{2}
photo-ionizing rate is small. We will further discuss these two models
in  \S \ref{sec:model}. In any case,  we are  led to the conclusion
that the shape  of the UV background spectrum  at a given redshift can
be quite different from the `canonical' UV background dominated by the
quasar population.

In  addition, the large  opacity seen in  several  regions of the STIS
spectrum of \objectname[]{HE 2347-4342} contrasts with the low opacity
$\tau_\mathrm{HeII}  =  1.88$ seen at   the same redshift (region $S$)
towards \objectname[]{Q 0302-003}. Despite the fact that STIS spectrum
of \objectname[]{Q   0302-003} has a    lower resolution, opacity gaps
similar to   the  ones   seen  in  the  spectrum of   \objectname[]{HE
  2347-4342} would have  been easily  detected.  Therefore, comparison
between these  two lines   of sight  confirm  our conclusion  that the
softness  of the UV  background shows  large  fluctuations at  $z \sim
2.8$.

\subsection{Metal-Line Ratios: 
Corroborating Evidence for a Soft Ionizing Spectrum}
\label{sec:high_ion}

We   have  shown   in the   previous  section  that     regions of the
\objectname[]{HE~2347$-$4342}  spectrum  in  which  the    \ion{He}{2}
opacity  is very high and the  corresponding \ion{H}{1} opacity is low
appear to  require photoionization by a very   soft UV radiation field
(see Figure \ref{fig:softness_vs_z}).  Independent evidence in support
of this conclusion is provided by the column  density ratios of highly
ionized  metals  observed at  high redshifts.   As several groups have
shown     \citep[e.g.,][]{GS97,SCD+97,Songaila98,KTA+99},    the
$N_\mathrm{SiIV}/N_\mathrm{CIV}$                                versus
$N_\mathrm{CII}/N_\mathrm{CIV}$  trends observed in  many high$-z$ QSO
absorbers   are inconsistent with  photoionization by  the standard UV
background  due  to   QSOs \citep[e.g.,][]{HM96,Madau99}.  Instead,  a
softer  ionizing spectrum   is   required    to produce  the      high
\ion{Si}{4}/\ion{C}{4} ratios observed in many systems.  This could be
due to either  (1) an important contribution to  the  ionizing flux by
stellar sources, or (2) strong  absorption of the \ion{He}{2} ionizing
photons emitted by the  QSOs before they  are able to travel far  from
the QSO.   The  latter  scenario  could  be  due to   \textit{delayed}
reionization of  the IGM,  as  advocated by  \citet{Reimers97}  or due
$z_\mathrm{abs} \approx z_\mathrm{em}$ absorption systems, such as the
system for \objectname[]{HE~2347-4342} discussed in \S 3.2.

\begin{figure}
  \begin{center}
  \epsscale{0.8}
  \plotone{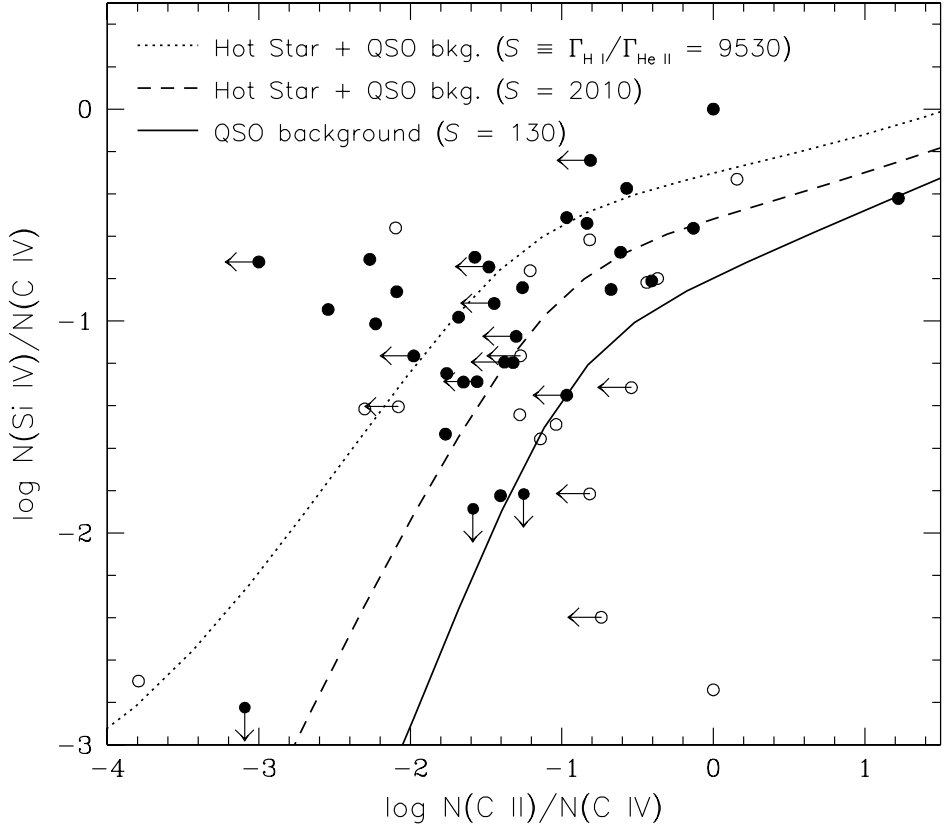}     
    \caption[]{Comparison of the \ion{Si}{4}/\ion{C}{4} versus
      \ion{C}{2}/\ion{C}{4} column density ratios observed in QSO
      absorption line systems at $z_{\rm abs} >$ 3.00 (filled circles)
      and $z_{\rm abs} <$ 3.00 (open circles) from \citet{Songaila98}
      to the ratios predicted by photoionization models for three
      different radiation fields. The radiation fields assumed for the
      three models have softness parameters $S = \Gamma _{\rm
        H~I}/\Gamma _{\rm He~II}$ ranging from $S$ = 130 to $S$ =
      9530. The ratios predicted by a photoionization model which
      assumes the UV background at $z$ = 3 due to QSOs {\it only} from
      the calculation of \citet{Fardal98} is shown with a solid line;
      this radiation field has $S$ = 130. The photoionization models
      shown with dashed and dotted lines assume the same QSO
      background plus a hot star (see text) which is 10 or 50 times
      brighter than the QSO background at the \ion{H}{1} Lyman limit.
      These models have $S$ = 2010 and 9530, respectively.
      \label{fig:highion}
      }
  \end{center}
\end{figure}

To demonstrate this  corroborating    evidence for a  soft    ionizing
radiation      field,     Figure~\ref{fig:highion}      compares   the
\ion{Si}{4}/\ion{C}{4}  versus  \ion{C}{2}/\ion{C}{4}  column  density
ratios observed at  high redshift by  \citet{Songaila98} to the trends
calculated    with   the photoionization   code CLOUDY  \citep[v90.04,
][]{FKV+98}   for three   ionizing   radiation  fields with   softness
parameters   ranging  from      $S =    130$ to    $S     =  9530$.    
Figure~\ref{fig:highion}  is similar to Fig.~17 in \citet{Songaila98},
except that we have also included systems with $\log{N_\mathrm{CIV}} >
14$.  For  the photoionization   calculations, we  adopted   the usual
plane-parallel geometry with constant density and set $N_\mathrm{HI} =
10^{15.3}~ \mbox{cm}^{-2}$  with a metallicity  of 1/100 solar and $[
\mbox{Si}/\mbox{C}]  \equiv           \log(\mbox{Si}/\mbox{C})      -
\log(\mbox{Si}/\mbox{C})_{\odot} = 0.0$.  In this figure, the observed
ratios  (and limits in  cases where \ion{Si}{4}  or \ion{C}{2} are not
detected)  are  shown with circles,  and the  filled circles represent
absorption systems at $z_\mathrm{abs}  >$  3.00 while measurements  at
$z_\mathrm{abs} <$ 3.00 are plotted with open circles. The solid curve
in Figure~\ref{fig:highion} shows  the  prediction resulting  from the
\citet{Fardal98}   UV  background  due   to QSOs   \textit{only}; this
radiation field has $S=  130$.    This ionizing spectrum    adequately
explains a  subset of  the  observed data.  However,  in  many of  the
absorption   systems    from     \citet{Songaila98},  the     observed
\ion{Si}{4}/\ion{C}{4}   ratios  are  substantially  higher   than the
prediction  of the photoionization model with  the  Fardal et al.  QSO
background. Furthermore, the discrepancies are  so large that it seems
unreasonable  to  attribute the difference   to an overabundance of Si
relative to C (see  below).  Instead, a substantially softer  ionizing
continuum appears to be required.

To illustrate  the  softness  of the   ionizing spectrum  required  to
reproduce  the  \ion{Si}{4}/\ion{C}{4} ratios,   the dashed and dotted
curves in Figure~\ref{fig:highion} show  the ratios predicted assuming
the \citet{Fardal98} QSO background  plus the  flux  emitted by  a hot
star according  to   a \citet{K91} Atlas   model  atmosphere; we  have
assumed a hot star with $T_\mathrm{eff}$ = 50,000  K which is 10 times
brighter (dashed line) or 50 times brighter (dotted line) than the QSO
background   at 1 Rydberg, but    have a negligible \ion{He}{2}  Lyman
continuum  flux.  These  radiation fields have  $S =  2010$  and 9530,
respectively.  These models  are intended for purposes of illustration
and are not expected to  be particularly realistic. A real ``stellar''
ionizing flux  source will obviously be  more complicated due to a mix
of  hot stars which contribute  to  the  continuum emission,  internal
reddening  and radiation transfer   effects.  Since the ionizing  flux
emerging from a starburst galaxy is poorly known at  this time, a more
sophisticated     model      does      not    seem    warranted.  From
Figure~\ref{fig:highion}  we see   that  a softer   ionizing continuum
provides better agreement with the observed  ion ratios. In many cases
the  radiation  field evidently must   be dramatically softer than the
QSO-only background.  This result is in agreement with our analysis of
the \ion{He}{2} GP absorption. A possible  explanation is the presence
of   young  starbursting galaxies close to the line of sight to the
QSOs used for this analysis. Such  nearby galaxies could  also provide
the metal enrichment  of the gas required  to explain detection  of Si
and C absorption lines in the first place.

It is   possible that  the   high  \ion{Si}{4}/\ion{C}{4}  ratios   in
Figure~\ref{fig:highion}    are partially   due  to overabundances  of
silicon relative to carbon \citep{SC96}. Silicon is an $\alpha -$group
element which  is rapidly produced in   short-lived massive stars that
undergo  Type II supernova explosions. This  is thought to produce the
well-known overabundance of Si relative to Fe in galactic stars of low
metallicity.    However,  the    nucleosynthesis  of  carbon   is more
complicated,  and   observed   trends   of   carbon   abundance versus
metallicity  are  highly uncertain in Milky  Way  stars \citep{McW97}. 
Models   of galactic  chemical   evolution  \citep[e.g.][]{TWW95}  can
produce  moderate ($3\times$)  overabundances  of Si   relative  to C,
[Si/C]  $\lesssim$   0.5.   Such overabundances   are  insufficient to
salvage  the   model      in    which  the  absorbers    with     high
\ion{Si}{4}/\ion{C}{4} ratios are  photoionized by  the QSO background
only, i.e., a softer background is  still required.  However, allowing
for  a moderate Si  overabundance enables a radiation  field with $S \ 
\approx$  2000   to  explain    a   large   fraction  of    the   high
\ion{Si}{4}/\ion{C}{4} ratios shown in Figure~\ref{fig:highion}.

A  cautionary note on metal  ratios  is of  order  here.  Based on the
analysis of 29 complexes,  \citet{Songaila98} also  presented evidence
for  a sudden change  in the \ion{Si}{4}/\ion{C}{4} ratio  at $ z \sim
3$,  in  the   sense that   the  median  value   of   this   ratio  is
$0.039_{-0.006}^{+0.009}$ for $z <  3$ and $0.13\pm0.04$  for $z > 3$. 
This observation is interpreted as  a change in the softness parameter
of the UV background consistent with the appearance of opacity gaps at
$ z \sim 2.9$ as observed by \citet{Reimers97} and in this paper.  One
can therefore wonder  if the \ion{Si}{4}/\ion{C}{4} ratios for systems
seen in the  spectra of \objectname[]{Q 0302-003} and \objectname[]{HE
  2347-4342}  also   show such a  trend   and are consistent  with the
softness derived from the observation of the \ion{He}{2} Gunn-Peterson
effect.  Table 2 from \citet{Songaila98} only lists systems outside of
the  proximity   zone,  at  more  than  $4000~\mbox{km}~\mbox{s}^{-1}$
blueward of  the  QSO  redshift.   For  \objectname[]{Q 0302-003}  and
\objectname[]{HE 2347-4342}, the highest redshift is therefore $3.04$.
The \ion{Si}{4}/\ion{C}{4} ratio is found  to be surprisingly constant
with redshift for these systems, with a mean value of $0.034\pm0.014$,
i.e.   unexpectedly close to the  median value found for $  z  < 3$ by
\citet{Songaila98}.  In  Figure~\ref{fig:highion}, these  points   are
located          within       the           ranges   $-1.3           <
\log{(N_\mathrm{SiIV}/N_\mathrm{CIV})}    <   -1.8$   and $-1.6      <
\log{(N_\mathrm{CII}/N_\mathrm{CIV})}   <   -0.5$  when \ion{C}{2}  is
observable,  i.e.  they  do  not require the  largest  values  for the
softness parameter.  However, in \citet{Heap00} and  in this paper, we
found a probable  association between \ion{C}{4} absorbers and opacity
gaps.   Therefore, if the  change  in the \ion{Si}{4}/\ion{C}{4} ratio
indeed corresponds to the appearance of opacity gaps, this association
may indicate that the  \ion{Si}{4}/\ion{C}{4} ratio does not  allow to
determine the mean softness of the UV background.  Instead, this ratio
would probe regions close to \ion{He}{2} opacity gaps, and therefore a
harder UV background. It is worth  noting that the apparent suddenness
of the  change in the \ion{Si}{4}/\ion{C}{4} ratio  at $ z  \sim 3$ is
not   confirmed in new  studies   by \citet{BSR01}  and \citet{KCD01}. 
Instead, these authors  find that this ratio  is remarkably flat  over
the range $2.0 < z < 3.5$ and $\bar{z} = 2.1, 3.3$ and 3.8,
respectively. They suggest that the \ion{Si}{4}/\ion{C}{4} ratio
alone may not be a good tool for the study of the \ion{He}{2}
re-ionization, while the UV background might be strongly affected by
galaxies at $z \sim 3$.

\section{Modeling the \ion{He}{2} Ly-$\alpha$ spectrum}
\label{sec:model}

In order to better understand   the causes of the opacity   variations
observed in the  \ion{He}{2} Ly-$\alpha$ spectrum of  \objectname[]{HE
~2347-4342}, we present  two models of the \ion{He}{2}  Gunn-Peterson
trough.  The   first  model (\S  \ref{sec:simple_model})  is  a simple
description of the opacity  gaps, aimed at understanding the relations
between the shape and  the maximal value of  the transmitted  flux, on
one hand, and the distance and the luminosity of a source close to the
line of sight thought to be responsible for the opacity gap.  However,
it  does not  solve  the transfer  equation   and  can only  serve  as
an illustration.  This failure is partially solved in the second
model  (\S \ref{sec:reminder}) which  is  an  expansion of  the  model
developed  in \citet{Heap00} to account for  multiple sources close to
the line of sight to the background quasar.

\subsection{Ionization State of the Intergalactic Medium}
\label{sec:model_ionization}

\subsubsection{Ionization state of the IGM}

At   large distances   from any   significant  ionizing sources,   the
photoionization  rates  are    mainly  due  to   the    UV background,
$\Gamma_i^\mathrm{J}$. In  this case,  Eq. \ref{eq:eta_44}  reduces to
$\eta = 0.44  ~ S^\mathrm{J}$, where we define  the softness of the UV
background        $S^\mathrm{J}$      as      $S^\mathrm{J}          =
\Gamma_\mathrm{HI}^\mathrm{J}/\Gamma_\mathrm{HeII}^\mathrm{J}     $.   
Observationally, it  is often impossible to distinguish $S^\mathrm{J}$
from  $S$. In other words, it  is often impossible  to know if a given
cloud  is photoionized   by a  diffuse  UV background   only, or by  a
combination  of a  diffuse UV background   and one or several discrete
sources.  However,   as  we will  see, opacity   gaps offer a probable
exception.

In the presence of additional ionizing sources, the photoionization
rates of a cloud $\Gamma_i$ can be written as the sums of the
photoionization rates due to the UV background and the photoionization
rates due to each individual source $s$:
\begin{equation}
  \label{eq:gamma_sum}
  \Gamma_i = 
\Gamma_i^\mathrm{J} 
+ 
\sum_{s = 1}^{n_\mathrm{s}} \, \Gamma_i^\mathrm{s}, 
\end{equation}
where $n_\mathrm{s}$ is the number of ionizing sources to consider.

In the absence of intervening absorption, the flux  at the Lyman limit
received by a cloud at redshift $z$ located on the  same line of sight
as the source  at redshift $z_s$  is related to  the source luminosity
$L_i^{s}$ by:
\begin{equation}
  \label{eq:inverse_square_law}
  f_i^{s}(z)   =  r(z,z_s)~ \frac{L_i^{s}}{4 \pi  D_\mathrm{L}^2 },
\end{equation}  
where $D_\mathrm{L}(z,z_{s})$ is the  luminosity distance between  the
cloud  and the    source $s$  \citep{KHP97}.  The   additional  factor
$r(z,z_s)$ takes into  account the relative time  dilation  as a Lyman
limit photon travels from the source whose redshift is now measured to
be $z_{s}$  to the given cloud whose   redshift is now  measured to be
$z$. One can show that $r(z,z_{s})$ is simply given by
\begin{eqnarray}
  \label{eq:r_z_zs}
 r(z,z_s) & = & \frac{1+z}{1+z_{s}}, ~~\mbox{if $z_s > z$, and}\nonumber\\
          & = & \frac{1+z_s}{1+z}, ~~ \mbox{otherwise}. 
\end{eqnarray}
However, for all the cases considered below, $r \simeq 1$ to within  4\%.

The luminosity
distance $D_\mathrm{L}(z,z_s)$  is related to the comoving distance
$D_\mathrm{C}(z,z_s)$ by $D_\mathrm{L} = D_\mathrm{C}/(1+z)$, if we
assume that the cloud at redshift $z$ and the source at redshift $z_s$
are both located along the same line of sight, and that $z \simeq z_s$.

Consider an ionizing source $s$  located at an
impact parameter $p_{\mathrm L}$ from the line of sight to the
background quasar. The luminosity distance that enters
the relation Eq.  \ref{eq:inverse_square_law} is given by:
\begin{equation}
  \label{eq:dl_sidesource}
  D_\mathrm{L}^2 =  D_{\mathrm{L},\parallel}^2+(p_\mathrm{L})^2,
\end{equation}
where $D_{\mathrm{L},\parallel}(z,z_s)$ measures the luminosity
distance along the line of sight to the background quasar between the 
cloud at redshift $z$ and the point on the line of sight closest to the
source $s$.

\subsection{A simple model for the opacity gaps}
\label{sec:simple_model}

In the  following, we present  a simple photoionization model which is
able to  reproduce   the shapes of  the  opacity gaps seen  in
existing  spectra of the \ion{He}{2}  Gunn-Peterson effect when the
effect of individual absorption lines are not too important.  We should
note that the  model presented here  is very similar in  principle to
the model of the proximity effect  developed by \citet{zd95} and also
suffers the same limitations.

We first note that $\tau_\mathrm{HeII} \propto N_\mathrm{HeII}$. Since
$N_\mathrm{HeII}          \propto     n_\mathrm{HeII}          \propto
(\Gamma_\mathrm{HeII})^{-1}$, one can easily show that
\begin{equation}
  \label{eq:tau_taumax}
  \tau_\mathrm{HeII} = 
  \tau_\mathrm{max}
  \left(
  1+\frac{
    \Gamma_\mathrm{HeII}^s
    }
    {
    \Gamma_\mathrm{HeII}^\mathrm{J}
    }
  \right)^{-1},
\end{equation}
where $\tau_\mathrm{max}$ is the  \ion{He}{2} optical depth  far
from a \ion{He}{2} ionizing source, i.e. one that arises from just
$\Gamma_\mathrm{HeII}^\mathrm{J}$. 

\begin{figure}[ht]
  \begin{center}
  \epsscale{0.6}
  \plotone{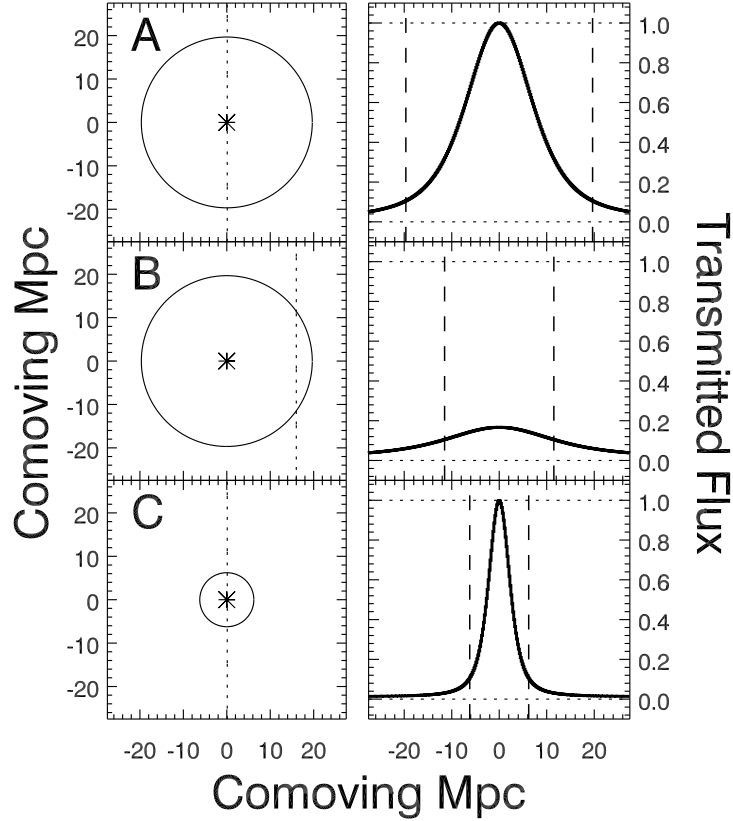} 
    \caption[]{Simple model for the opacity gaps.  
      \textit{Left panels:} Schematic plan view of the different
      examples discussed in the text.  
      The stars represent the `hard' sources
      S1 or S2, of luminosities $L^\mathrm{S1} = 2.5 \times
      10^{29}~\mbox{ergs}~\mbox{s}^{-1}~\mbox{Hz}^{-1}$  (panel
      \textit{(A)} and \textit{(B)})    and       $L^\mathrm{S2} =
      L^\mathrm{S1}/10$ (\textit{panel (C)}).
      The solid circles indicate
      where the photoionization rate due to the source S$i$
      $\Gamma_\mathrm{HeII}^{\mathrm{S}i}$ equals the one due to the UV
      background assumed to be $\Gamma_\mathrm{HeII}^\mathrm{J} = 3.75
      \times
      10^{-15}~\mbox{ergs}~\mbox{s}^{-1}~\mbox{Hz}^{-1}~\mbox{sr}^{-1}$.
      The dotted lines  represent the 
      lines of sight passing close to the source S$i$.  
      \textit{Right panels:} Corresponding shapes for the opacity
      gaps. The  dashed lines indicate where 
      $\Gamma_\mathrm{HeII}^{\mathrm{S}i} =
      \Gamma_\mathrm{HeII}^\mathrm{J}$, leading to $\tau_\mathrm{HeII} =
      \tau_\mathrm{max}/2$ (i.e., $\tau_\mathrm{HeII} = 2.25$ in these
      examples).       
    \label{fig:simple_opacitygap_examples}
}
\end{center}
\end{figure}

In    such     a   simple     model,  the      photoionization    rate
$\Gamma_\mathrm{HeII}^s$  only depends on  the distance $D_\mathrm{L}$
and the source luminosity and spectral shape  at the \ion{He}{2} Lyman
limit (i.e., the flux  of  \ion{He}{2} ionizing photons).   Therefore,
for a given value of the opacity far from the ionizing source $s$, the
shape of the  opacity gap is  mainly determined by two quantities: the
luminosity of  the   source  at the  \ion{He}{2}   Lyman  limit  (more
precisely, the flux of \ion{He}{2} ionizing photons), which determines
the overall extent of the gap, and its distance from the line of sight
$p_\mathrm{L}$, which  determines how  `flat' the  gap appears  in the
spectrum.       This      point      is  illustrated      in    Figure
\ref{fig:simple_opacitygap_examples}. For $\tau_\mathrm{max}$, we use
the value 4.5, which is the limit on the opacity in region $D$. The
results do not change much for sightly larger (4.8, as in region $C$)
or slightly smaller (3.7, as in region $I$) values. Much larger values
of $\tau_\mathrm{max}$ require proportionally larger luminosities.
In panels
(A) and (B), a source S1  of luminosity at the \ion{He}{2} Lyman-limit
$L^\mathrm{S1}            =        2.5          \times         10^{29}
~\mbox{ergs}~\mbox{s}^{-1}~\mbox{Hz}^{-1}$  and   a flat   spectrum in
$\lambda$ blueward of 4 Rd is represented by a star.  The limit of the
zone         over         which     its      photoionization      rate
$\Gamma_\mathrm{HeII}^\mathrm{S1}$  is  larger  than  the  \ion{He}{2}
photoionization      rate  of the    UV background,     assumed to  be
$\Gamma_\mathrm{HeII}^\mathrm{J}   =       3.75      \times   10^{-15}
\mbox{s}^{-1}$, is   represented by the  solid  circle.   The lines of
sight represented  by  dotted lines cross   that zone.  The  resulting
shape for the opacity gaps are represented  in the two upper graphs of
the panels  on the right.   The  vertical  dashed  lines indicate  the
distances  from    the   points    closest   to  the   sources   where
$\Gamma_\mathrm{HeII}^\mathrm{S1} =  \Gamma_\mathrm{HeII}^\mathrm{J}$,
i.e., $\tau_\mathrm{HeII} =  \tau_\mathrm{max}/2$.   The limit of  the
ionizing zone  for  a  second  source S2  10 times  less   luminous is
represented by  the  dotted  circle in   panel (C).  A  line of  sight
passing  close to  the center  leads to  the  opacity gap  seen in the
bottom graph of the right panel.

These examples show that the  overall scale of  the gap (measured, for
example, by the  full width at  half maximum) is mainly determined  by
the luminosity of  the source. On the  other hand, for  a given scale,
the amount of transmitted flux  at the center  of the opacity gap is a
direct measure  of how close  the line of sight passes  by the source. 
In particular, it is also clear from such graphs  that, because of the
presence of the UV   background radiation, an  opacity gap  should not
show  any  sharp boundary  as it   is the  case  for a Str{\"o}mgren's
sphere.

\begin{figure}[ht]
  \begin{center}
    \epsscale{1.0}
    \plottwo{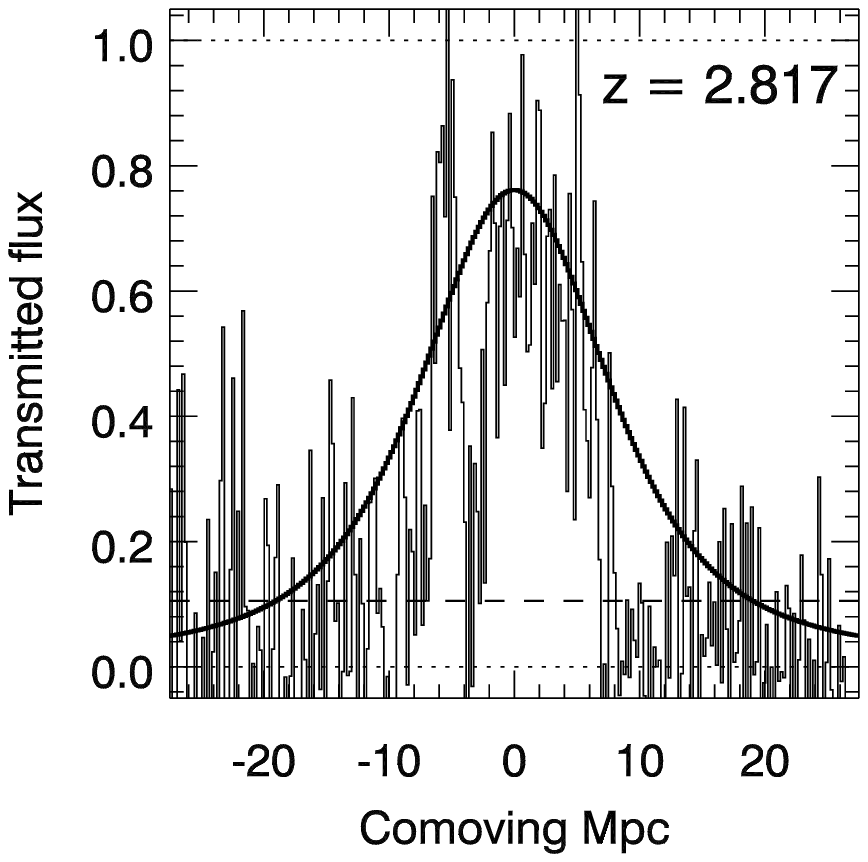}{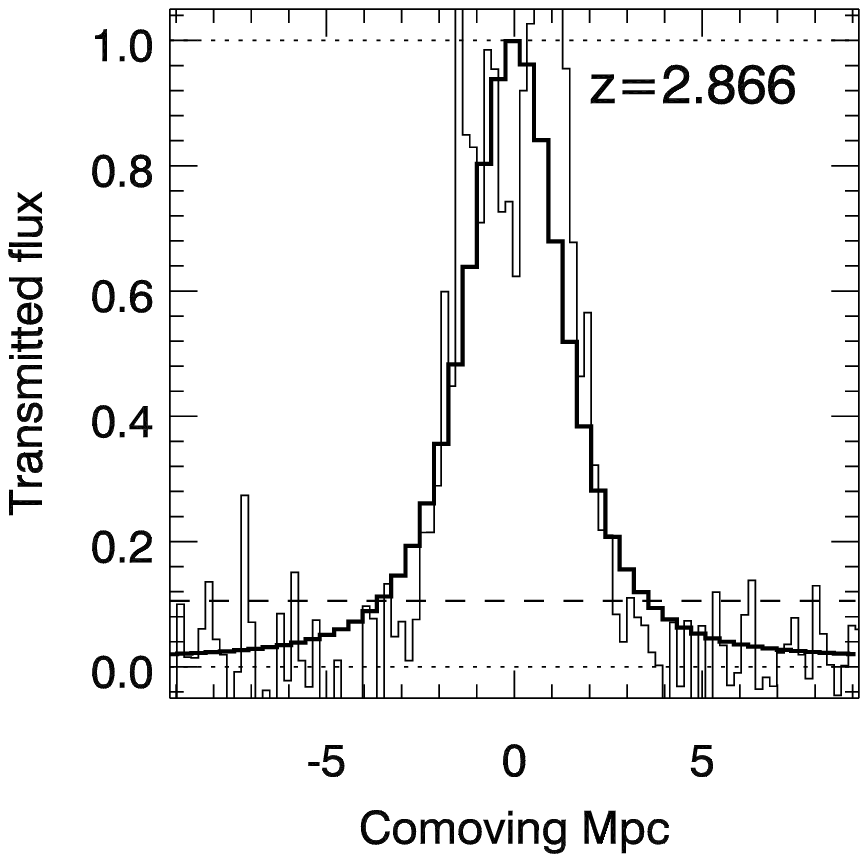} 
    \caption[]{Models for the observed opacity gaps.
      \textit{Left panel:}
      Model for the $z = 2.817$ gap.  
      \textit{Right panel:} Best attempt to model the $z = 2.866$
      gap.  Such a simple model of course fails to reproduce
      the effect of individual absorption lines which probably causes
      the sharpness of the profile.
    \label{fig:simple_opacitygap_data}
}
  \end{center}
\end{figure}

The left panel of Fig. \ref{fig:simple_opacitygap_data} shows that the
$z = 2.817$ opacity gap in the spectrum of \objectname[]{HE~2347-4342}
can be modeled by a line  of sight passing  at about 5.0 comoving Mpc
from a source  with a luminosity at  the \ion{He}{2} Lyman-limit  $L =
2.5 \times 10^{29} ~\mbox{ergs}~\mbox{s}^{-1}~\mbox{Hz}^{-1}$, similar
to the source $S1$ in the example above.  Similarly, the line of sight
(B)  in Fig.  \ref{fig:simple_opacitygap_examples} represents  well the
low opacity seen in  region  $S$ at $\bar{z}  =  2.82$ on the  line of
sight towards \objectname[]{Q~0302-003} \citep{Heap00}.

The right panel of Fig. \ref{fig:simple_opacitygap_data} is an attempt
to model the $z  =  2.866$ opacity  gap.  The  ionizing source  has  a
luminosity  at the  \ion{He}{2}  Lyman-limit   $L  = 10\time   10^{27}
~\mbox{ergs}~\mbox{s}^{-1}~\mbox{Hz}^{-1}$  and  is located  at  0.070
comoving Mpc from the line of sight. This  graph illustrates the limit
of  this  model as  it  does  not take into    account the presence of
\textit{known} intervening absorption  systems, and, consequently,  is
unable to  reproduce   the sharpness of the    opacity  gap.  In   the
following section,  we  try to  remedy   this problem by   solving the
cosmological equation  of transfer using as much  as possible  all the
observable  quantities,  in  particular,  the   observed   intervening
absorption systems.  In order to  model   the opacity gaps  and  other
features  seen  in  the spectrum  of   \objectname[]{HE 2347-4342}, we
expand the method presented in \citet{Heap00}  in order to account for
the presence of  additional sources of  radiation close to the line of
sight. In the next section, we present a brief description of the
method.

\subsection{Model of the \ion{He}{2} Gunn-Peterson effect}
\label{sec:reminder}

\subsubsection{One source: the first cloud}
In   \citet{Heap00},  we only considered one     source, which was the
background   QSO  \mbox{Q~0302--003}   itself.   In   this  case,  the
photoionization rates   for each cloud   can  be obtained recursively,
starting with the   first cloud $k$ close  to  the  QSO.   Indeed, the
photoionization rates of the first cloud at $z = z_\mathrm{1}$ are:
\begin{equation}
  \label{eq:1st_cloud}
  \Gamma_i^{\rm 1} = 
  \Gamma_i^\mathrm{J}
  +
  \frac{\sigma_i}{h} ~
    \frac{f_i^{\rm QSO}}{3+\alpha_i^{\rm QSO}}  ~
    \left(
      \frac{1+z_\mathrm{QSO}}{1+z_1} ~
    \right)^{-\alpha_i^{\rm QSO}},
\end{equation}
where the QSO flux at a redshift $z$ and frequency $\nu \ge \nu_i$
is $f^{\rm QSO}(z,\nu) = 
f_i^{\rm QSO}(z)~[(1+z_\mathrm{QSO})~\nu/\nu_i]^{-\alpha_i^{\rm QSO}}$.
The photoionization rates of \ion{H}{1} and \ion{He}{2} 
are calculated with Eq. \ref{eq:1st_cloud}, and then 
Eq. \ref{eq:eta_44} is used to compute $\eta$.  The value of the
\ion{He}{2} column density  is then easily calculated from the observed
\ion{H}{1} column density.

\subsubsection{One source: the $k > 1$st  cloud}  
The QSO spectrum below the ion $i$ Lyman limit seen by the $k$th
cloud ($k \ge 2$) is depressed by the continuum opacity produced by
the $k-1$ clouds located between it and the QSO. In \citet{Heap00}, we
obtained a series expansion for the QSO contribution to the
photoionization rate $\Gamma_i^k$ which can be written as
\begin{equation}
  \label{eq:l_cloud_analytical}
  \Gamma_i^k = 
  \frac{f_i^\mathrm{QSO}~\sigma_i}{3 ~ h}~
  \left(
    \frac{1+z_\mathrm{QSO}}{1+z_k}~
  \right)
  ^{-\alpha_i^\mathrm{QSO}}~
  \tau_\mathrm{eff}
  ^{
    -\frac{3+\alpha_i^\mathrm{QSO}}{3}
  }~
  \gamma^*\left(
    \frac{3+\alpha_i^\mathrm{QSO}}{3},\tau_\mathrm{eff}
  \right),
\end{equation}
where
$\tau_\mathrm{eff} =  \sum_{l = 1}^{k-1} ~
    ~ N_i^l ~\sigma_i~
    \left(
      \frac{1+z_k}{1+z_l} 
    \right)^3
$
and
$\gamma^*(a,x)$ is the incomplete Gamma function.

\subsubsection{Multiple sources on the line of sight}

We have just  shown   how  the  \ion{He}{2}  column density  can    be
estimated, based  on  the \ion{H}{1}  column density of  the  observed
clouds, on the  luminosity of the  background QSO and on the intensity
of the  UV    background. In  the   process, we  had to  calculate  in
particular    the  contribution  from the   background    QSO  to  the
photoionization rates at the redshift of each cloud.

\begin{figure}[ht]
  \begin{center}
  \plotone{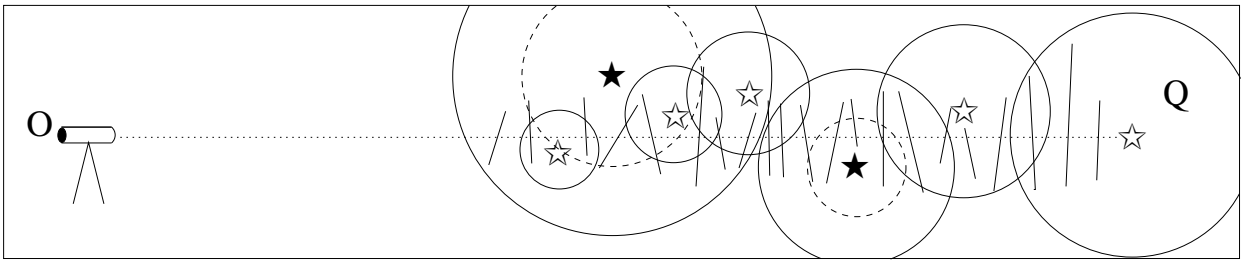} 
    \caption{Schematic representation of the distribution of
      clouds (thin lines)  and sources between  the observer ($O$) and
      the background quasar ($Q$).   `Hard'   and `soft' sources   are
      represented by filled and open stars, respectively.  The regions
      where the \ion{H}{1}  and \ion{He}{2}  photoionization rates due
      to each  source equal the UV background  ones are represented by
      the solid and dashed circles, respectively.  }
    \label{fig:scheme_model}
  \end{center}
\end{figure}

Taking into  account other  sources  on the  line of  sight  follows a
similar    procedure.  Fig.   \ref{fig:scheme_model}     schematically
represents the distribution of sources and absorbers along the line of
sight towards the background quasars.  However, whereas the values for
each $\Gamma_i$ (and consequently for  the column densities) could  be
obtained in one  pass in the presence of  one source, the presence  of
other sources distributed on or close to the line  of sight to the QSO
requires several calculations for each $\Gamma_i$ until convergence is
reached.

We found that the following algorithm converges after a few passes for
a small number of sources.  As before, (1) we compute the contribution
from the quasar to the $\Gamma_i$ of the first cloud: $\Gamma_i(z_1) =
\Gamma_i^\mathrm{J}+\Gamma_i^\mathrm{QSO}(z_1)$. (2) From the observed
$N_\mathrm{HI}$  of  the first cloud,  we  use Eq.  \ref{eq:eta_44} to
calculate $N_\mathrm{HeII}$ of the first  cloud.  (3) Successively, we
repeat  step (1) and  (2) using Eq.  \ref{eq:l_cloud_analytical} until
the contribution from  the QSO to the  values  of $\Gamma_i$  for each
cloud  of the  sample  has been   evaluated and, consequently  all the
$N_\mathrm{HeII}$. (4) Similarly, we  compute the contribution of  the
first source  (as in step  (1)) to the first  cloud at higher redshift
than the 1st  source, and  recalculate  the $N_\mathrm{HeII}$  of  the
first cloud, as in step (2). (5) Similarly as  step (3), we repeat the
procedure for all the clouds  at higher redshift than  the 1st source. 
(6) we proceed as in  (4)  and (5), but   for all the clouds at  lower
redshift than the first source. (7) We repeat steps (4) to (6) for all
the other  sources considered. (8)  Since  the presence  of additional
sources make the IGM more transparent, the radiation  from the QSO and
the other sources reaches further away than in  the case when only one
source  is considered; consequently, the   whole procedure from (1) to
(7) must be    repeated,  but  this  time  taking   into  account  the
contribution to $\Gamma_i$  from all the  sources as evaluated in each
previous pass.  (9) Convergence is  reached when the $\Gamma_i$ do not
vary significantly ($\la 1\%$) between two successive passes.

There  is a severe   limitation to  this method:   we must  assume, as
depicted  in Fig.    \ref{fig:scheme_model}, that  the clouds  located
between the line of sight and a given source  located at some distance
from the line of sight are the same  and have the same characteristics
($N_\mathrm{HI}, b$) as the ones located on the sight line to the QSO.
In particular, we must assume that none of them are optically thick.

Figure  \ref{fig:model_opacitygap}  presents   the  results   for  two
different models.   The top panel  shows the  calculated appearance of
the \ion{He}{2}  Gunn-Peterson trough if only two  AGN are  present on
the  line  of     sight  towards  \objectname[]{HE~2347-4342}.   Their
characteristics  are  listed   in    the  first  two lines  of   Table
\ref{tab:parameters} and  model the opacity gaps as  well as possible. 
Note that the 3rd and 4th  columns give the luminosities for isotropic
sources.  The values  are relatively well  constrained (within 50\%).  
The  following  two columns list  the   corresponding expected fluxes. 
They indicate that the sources responsible for the opacity gaps should
be easily visible in the optical bands if they are not hidden by dust:
the sources responsible for the $z = 2.815$ and 2.866 gaps should have
a magnitude $R  \simeq 18.9$ and  20.2, respectively,  if  they have a
flat spectrum in $f_\lambda$.  A better approach  to reveal them is to
search for X-ray sources with \textit{Chandra} or \textit{XMM-Newton}.
Compared   to     the   simple    model   presented    earlier     (\S
\ref{sec:simple_model}), the  fact that we  now take  into account the
presence of individual  absorption lines  allows  us to reproduce  the
sharpness of the $z  = 2.866$ opacity  gap. However, it is  clear that
such a model is unable to reproduce the  high opacity regions, such as
regions $C$, $D$ and $E$.  The model shown in the bottom panel of Fig.
\ref{fig:model_opacitygap}  seems to reproduce  the observed  spectrum
much better: its characteristics are discussed in the next section.

\subsubsection{Possible origins for the high opacity regions}

It  is worth  discussing the possible   origins  for the high  opacity
regions.  In this section, we  limit  ourselves to two scenarios
both compatible  with photo-ionization, although  not necessarily with
photo-ionization equilibrium.    In particular,  these two  models are
compatible with the   map of  the softness  parameter  shown in  Fig.  
\ref{fig:softness_vs_z}.  First,  we   recall the  \textit{delayed
  re-ionization}   model invoked  by   \citet{Reimers97}.   Then,   we
introduce  a  \textit{soft source}  model.   Possible alternatives  to
these two models will be discussed later (\S \ref{sec:alternative}).

The    \textit{delayed re-ionization} model \citep{Reimers97} suggests
that the spectral  ranges that show  a  large softness $S$  arise from
regions  of  the Universe   where  hydrogen has  been   already mostly
re-ionized but where the \ion{He}{2}  ion density $n_\mathrm{HeII}$ is
still larger  than  in regions  of mean  $S$.  The reduced \ion{He}{2}
ionizing photon flux causing the larger $n_\mathrm{HeII}$ results from
the fact   that \ion{He}{2} ionizing  photons  has not yet  managed to
doubly ionize  He in  the intergalactic  medium;  in other  words, the
regions of large $\tau_\mathrm{HeII}$  are   still shielded from   the
sources  of \ion{He}{2} ionizing  photons  due to the relatively large
$n_\mathrm{HeII}$  in the intergalactic   medium.  In this  model, the
\ion{H}{1} ionization rate  $\Gamma_\mathrm{HI}$ is typically constant
over the  redshift range of  interest, but the  \ion{He}{2} ionization
rate $\Gamma_\mathrm{HeII}$ is small.   An implicit hypothesis of this
model is that the sources of both  \ion{H}{1} and \ion{He}{2} ionizing
photons are distant from the regions of large $\tau_\mathrm{HeII}$.

Here,   we propose a second   scenario:  the presence of  \textit{soft
  sources} close to  the line-of-sight of the quasar  lead  to a local
increase of  \ion{H}{1} ionizing  photons without providing additional
\ion{He}{2} ionizing photons.   Such sources could  be  galaxies, e.g. 
Lyman-break  galaxies  which  have   recently  been found  to   have a
significant  flux below 1  Ryd  \citep{Steidel+00}.  They can also  be
quasars   similar to \objectname[]{HE   2347-4342},  i.e.   which have
$z_\mathrm{abs}   \simeq  z_\mathrm{em}$ absorption   systems that are
basically transparent to  \ion{H}{1}  ionizing photons but   opaque to
\ion{He}{2}  ones.   A third, although  less  likely, possibility is a
normal quasar close to  the line of sight  but located behind a  cloud
with  an  \ion{H}{1}  column  density low  enough  that  the  cloud is
optically     thin    for  \ion{H}{1}      ionizing   photons   (i.e.,
$\log{N_\mathrm{HI}} \la 17$)  but with an \ion{He}{2}  column density
large   enough to be  optically  thick to \ion{He}{2} ionizing photons
(i.e.,   $\log{N_\mathrm{HeII}}   \ga 18$).  In    this  scenario, the
intensity  of the  UV background  at  the \ion{He}{2}  Lyman  limit is
relatively  similar  in  regions of  mean   or  large  $S$,  as  it is
determined by (not  too) distant sources, while  the intensity  at the
\ion{H}{1} Lyman  limit  is increased by  a large   factor due  to the
presence of the soft sources close to the line-of-sight.

\begin{figure}[ht]
  \begin{center}
  \plotone{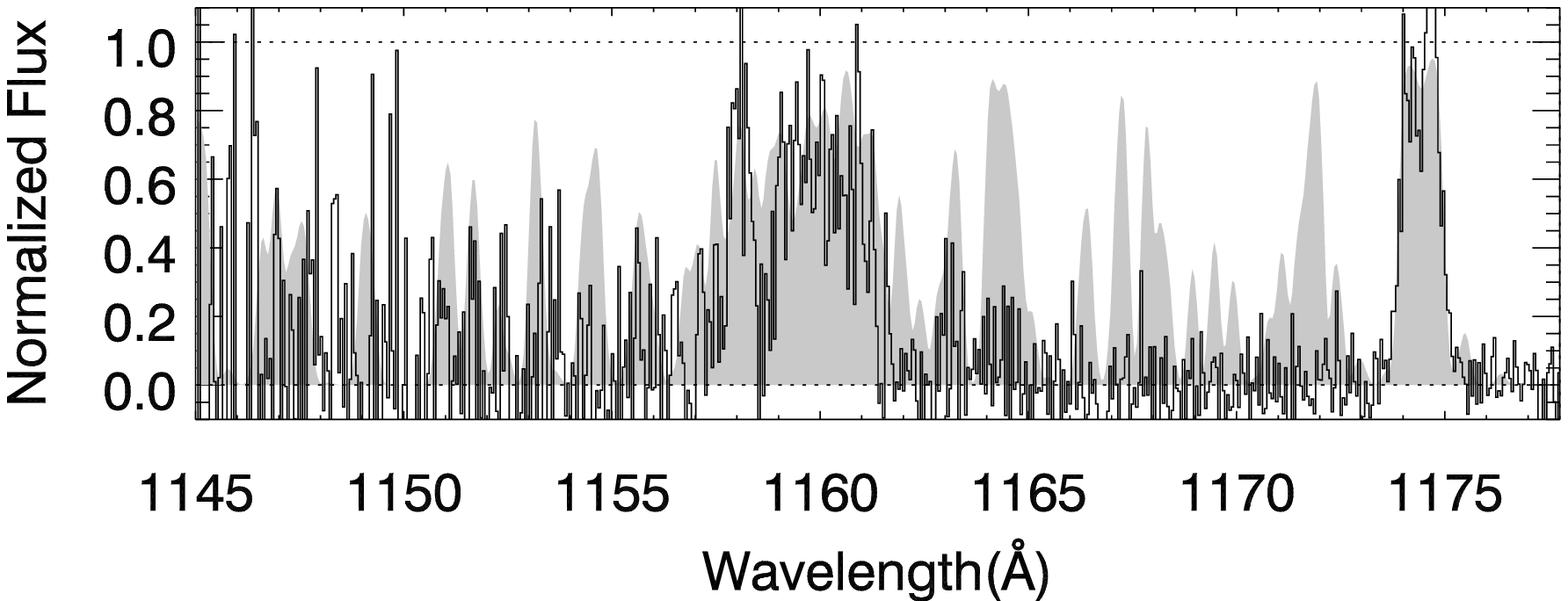} 
  \plotone{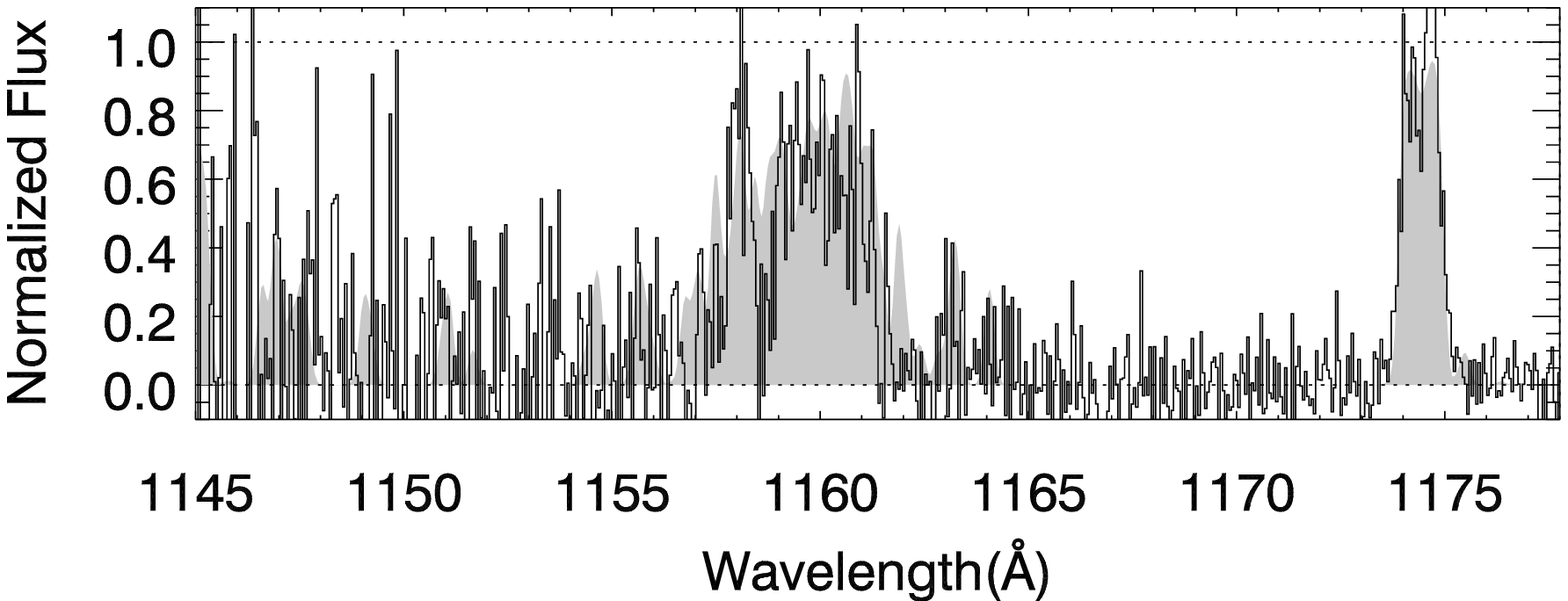} 
    \caption{Comparison of models of the \ion{He}{2} Gunn-Peterson
      trough (shaded)  to  observations (line).   Assuming only 2  AGN
      close to the line of sight to HE~2347-4342, the opacity gaps are
      well reproduced  but  not the  spectral  ranges of high  opacity
      \textit{(top panel)}. In addition to the  2 AGN, we have added 7
      `soft'  sources close  to the   line  of sight   \textit{(bottom
        panel)}. The  increased amount of \ion{H}{1} ionizing photons
      lead to a    larger $\Gamma_\mathrm{HI}$ and,   therefore, to  a
      larger softness parameter $S$.
    \label{fig:model_opacitygap}
    }
  \end{center}
\end{figure}

The bottom  panel of Fig.  \ref{fig:model_opacitygap} illustrates  how
the introduction of a small  number of soft  sources can reproduce the
observed spectrum of \objectname[]{HE 2347-4342}. The only constraints
imposed on this  model is that the mean  opacities in each regions are
broadly  met and, in  particular, that  no  peak  of transmitted  flux
appear in regions   of high  opacity.   The  solution shown here    is
certainly not  unique: we have no way  to determine even the number of
sources involved.  In   particular, such a  lack of  constraints means
that we may as well choose a small number of bright sources as a large
number  of faint  ones.   In addition,  since the  lines composing the
diffuse  component  are not observed   but  have their characteristics
randomly  selected, different realizations  lead to slight differences
in the appearance  of  the simulated \ion{He}{2} Gunn-Peterson  trough
even for  identical model parameters.   For  illustrative purposes, we
list  in Table \ref{tab:parameters}     the parameters used  for   the
solution shown in the  Fig.  \ref{fig:model_opacitygap} (bottom  panel). 
The hard sources were discussed in the previous section.  On the other
hand, we see that the `soft' sources should be revealed in deep images
of the    field.   In the  model described   above,  the   sources are
relatively  bright galaxies and should  be  easily detectable ($R \sim
20.6$ to  $21.4$).  In this case,  their local space density  is $\sim
10^{-1}~(\mbox{comoving ~  Mpc})^{-3}$, a  factor $\sim   10^3$ larger
than for Lyman break  galaxies \citep[LBG,  ][]{Steidel+96}.  However,
the  important quantity is  the number density of \ion{He}{2} ionizing
photons  or luminosity density.  Therefore,  the  sources can be  more
numerous and fainter and may  indicate the presence of a proto-cluster
close to  the line of sight towards  \objectname[]{HE 2347-4342}.  The
steepness of the  luminosity  function of the LBG   \citep{Steidel+99}
actually insures that the required  luminosity density can be obtained
with  a space   density  of  galaxies comparable  to   the  LBG one.   
\citet{Steidel+00} showed that their contribution to the UV background
intensity is important and can be predominant compared to the QSO one,
but still be compatible with the value  of the UV background intensity
at the \ion{H}{1} Lyman limit as measured by the proximity effect.  It
is also  possible that  most of  the  \ion{H}{1} ionizing photons  are
produced by one AGN,  as bright at 1  Rd as the  one causing the $z  =
2.817$ opacity gap, but with a negligible flux of \ion{He}{2} ionizing
photons.  Despite the uncertainty in  the mix of ionizing sources, the
important point  is that photoionization  alone allows us to reproduce
the observations.

\begin{figure}[ht]
  \begin{center}
  \plotone{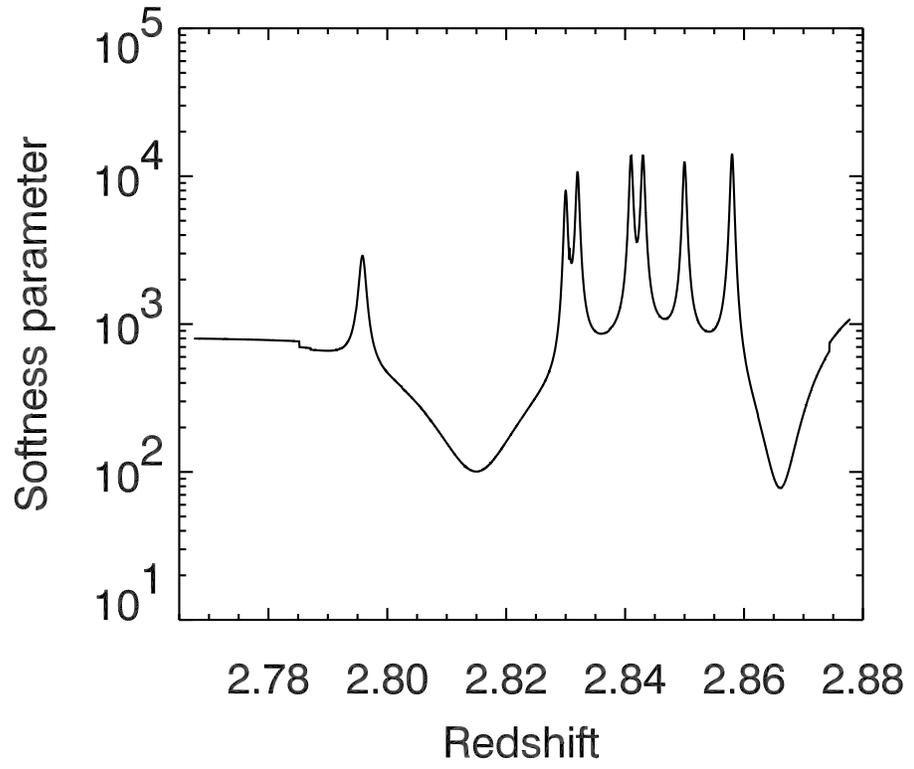}     
        \caption{Softness parameter as a function of redshift for the
          model considered in Fig. \ref{fig:model_opacitygap}, bottom panel.
    \label{fig:softness_model}
    }
  \end{center}
\end{figure}

Figure   \ref{fig:softness_model}   shows the   map   of the  softness
parameter corresponding to the model shown at the bottom panel of Fig.
\ref{fig:model_opacitygap}. Interestingly, the softness reaches values
close  to $S \sim 1 \times  10^{4}$  at its peaks,   where the line of
sight  is  closer to the  sources.  If the impact  parameters are even
smaller than considered here, one can expect  even larger value of $S$
as  well  as  formation of absorption   lines, such  as \ion{C}{4} and
\ion{Si}{4}.  Therefore,  the situation  presented  here is consistent
with the large values of the  \ion{Si}{4} to \ion{C}{4} column density
ratios seen    in quasar absorption lines   which  we  examined  in \S
\ref{sec:high_ion}.

Of course, it may very well be that the two scenarios could co-exist
even over a small redshift range.  But can existing observations allow
us to discriminate between the two models?  We first note that in the
delayed ionization model, the \ion{H}{1} lines should have the same
mean number density and column density distribution in spectral ranges
of mean or large $S$ (i.e., opacity gaps excluded).  However, in
regions where \ion{He}{2} re-ionization is delayed, the temperature of
the gas is expected to be lower and therefore the Doppler
$b$-parameter of the lines should be smaller than in regions where the
\ion{He}{2} re-ionization has taken place \citet{STRES00}.  

Since the line  broadening is mostly turbulent,  the best way to  test
this hypothesis would be to compare the cut-off $b_\mathrm{c}$ seen in
the $b$ distribution \citep[e.g.,][]{KHC+97}  of the \ion{H}{1} lines:
we would  expect  $b_\mathrm{c}$ to be  smaller  for  the  lines whose
corresponding \ion{He}{2}  lines fall in the  regions $C$ and $D$ than
for the rest of the spectrum.
However, a reliable  determination of $b_\mathrm{c}$  requires a large
number of lines in the  sample, $\sim 200$ \citep{STLE99}, while  only
15  are  detected in  regions $C$  and  $D$.   Therefore,  we can only
compare  the mean $\bar{b}$ and  standard deviation $\sigma(b)$ of the
$b$ parameters for  the two samples. We find  them to be similar, with
$\bar{b}_\mathrm{C,D} = 32\pm13~\mbox{km}~\mbox{s}^{-1}$ compared   to
$\bar{b}_\mathrm{other}   =   34\pm17~\mbox{km}~\mbox{s}^{-1}$.     In
addition, a 2-sided Kolmogorov-Smirnov test   of  the samples of   $b$
values gives a D statistic of 0.22 with a probability  of 0.53 that it
arises by chance.   In other words,  the two samples appear drawn from
the  same population.  Unfortunately, the small  number  of lines used
for the test makes it of limited use.

On the other hand, in the soft  sources model, one expects an increase
of the \ion{H}{1} ionization fraction of the gas and therefore a local
decrease  in the number density  and  column density of the \ion{H}{1}
lines.  Comparison between the number of lines observed in regions $C$
and $D$ and the number of lines  expected based on the observations by
\citet{KHC+97} show agreements for  different column density  cut-off. 
We note, however, that the scatter in the  number of expected lines is
itself much larger  than if the  error were purely  Poissonian so that
this test is not very sensitive over a small spectral range.  However,
it is worth noting  that the region   of low \ion{H}{1} opacity  at $z
\sim   2.855$   is quite   remarkable   for   its  extent: we  measure
$\tau_\mathrm{HI} \simeq 0.0085$ over   a spectral range larger   than
$\Delta  \lambda =  6.5$\AA\,  (of course,  such  a measure is largely
affected  by  errors on    the  continuum definition).   There   is no
occurrence of such  a large spectral range   with such low  \ion{H}{1}
opacity in the spectrum  of  \objectname[]{HE 2347-4342} before  $z  =
2.6$. Another way to express this observation is that there is no line
with  $N_\mathrm{HI}  >  3  \times   10^{12}~\mathrm{cm}^{-2}$ over this
spectral range.  From Table 4 of \citet{KHC+97}, the \ion{H}{1} column
density    distribution  at a mean    redshift   $\bar{z}  = 2.85$  is
$f(N_\mathrm{HI})  =   4.9   \times  10^{7}  ~N_\mathrm{HI}^{-1.46}$.  
Therefore, we  expect to see $\sim  760$ lines with $N_\mathrm{HI} > 3
\times 10^{12}~\mathrm{cm}^{-2}$  per  unit redshift, or about  4 such
lines over a spectral  range of 6.5\AA\,  at $z  \simeq 2.885$. For  a
Poisson distribution, the \textit{a   posteriori} probability that  no
such lines are seen is  $< 0.02$.  An  additional argument in favor of
the soft sources model come from the recent study by \citet{BSR01}: by
comparing  the \ion{Si}{4}/\ion{C}{4}     vs the \ion{C}{2}/\ion{C}{4}
ratio  over the range $1.9  < z <  4.4$ with  CLOUDY models, they find
that the data are best explained by a UV  background dominated by QSOs
at   low  redshift  but  that   a dominating   stellar contribution is
progressively required at higher redshift.

Finally, we should  note that in the absence  of soft sources,  within
the  delayed reionization model, the  high opacity in regions $A$, $C$
and $D$ puts constraints either on the  the lifetime or the luminosity
of the AGN causing the opacity gaps.  Indeed,  let us suppose that the
opacity   gap  at  $z  = 2.86$  is  indeed   caused  by  an  AGN whose
characteristics    are  close    to   the values    given   in   Table
\ref{tab:parameters}.   Therefore,  the large opacity  in regions $A$,
$C$ and $D$ indicates at least one of the two following conclusions:

\textit{(i)}  Let us  assume that there   are  no high  column density
systems  (including  systems intrinsic to the   AGN) on the light path
between the AGN (located off the line-of-sight) and regions $A$ to $D$
and that the  time that was  required to photo-ionize the region  $B$,
$t_\mathrm{phot}$, is similar to the time required to photo-ionize the
regions  $A$, $C$ and  $D$.   Therefore, the  difference  between  the
lifetime of the AGN, $t_\mathrm{AGN}$, and $t_\mathrm{phot}$ is larger
than the time required for light to go from  the AGN to region $B$ but
smaller than the   time required to go from   the  AGN to any of   the
regions $A$, $C$ and,  \textit{a fortiori}, $D$. Consequently, we have
$10^6~\mathrm{yr} \la t_\mathrm{AGN}  - t_\mathrm{phot} \la  10 \times
10^6~\mathrm{yr}$, compatible with currently admitted AGN lifetime.

\textit{(ii)}  In unification scheme  models, the AGN continuum region
may  be  surrounded by  a molecular  torus   optically  thick to  both
\ion{H}{1} and \ion{He}{2} ionizing photons.  In  this case, the model
that  we have developed   would require  to   take into   account more
variables.  Indeed, the  size and shape  of the opacity  gaps will not
only be determined by  the luminosities and  impact parameters  of the
responsible AGN but also by the orientation  of the ionizing radiation
cone  relative  to   the   line  of sight   towards   \objectname[]{HE
  2347-4342}.   Of course, the possible   configurations are numerous. 
However, it  is   interesting  to consider the   possibility  that the
ionizing  cones are  oriented  perpendicularly  to the   line of sight
towards \objectname[]{HE  2347-4342}  as this configuration  allows to
derive lower limits to the AGN luminosities.  In  the case of the $z =
2.866$  opacity gap, region $B$  alone would  see the continuum region
while the torus blocks ionizing photons to reach  regions $A$, $C$ and
$D$, and therefore would explain their high \ion{He}{2} opacity in the
absence of other hard sources.  Since region $B$ extends over $\sim 3$
comoving Mpc,  and  assuming a half-opening  angle  of  $\sim 45$  deg
\citep[][]{B89}, we derive an impact  parameter of $\sim 1.5$ comoving
Mpc for the AGN.  Since this value is only slightly larger (50\%) than
the impact parameter we found above (cf.  Table \ref{tab:parameters}),
the luminosity  of the AGN  should  be at least of   the same order of
magnitude as listed in Table \ref{tab:parameters}.  We can do the same
exercise for  the $z  = 2.817$  opacity   gap.  The comoving  distance
between its extremities  (broadly  taken  to be from   1157.75\AA\, to
1162\AA,   although residual signals  are   detected at the  3$\sigma$
levels in regions $E$ and $H$) is  $\sim 15$ comoving Mpc.  Therefore,
the impact parameter would be $\sim 7.5$ comoving Mpc, also about 50\%
larger    than  the value   listed    in  Table \ref{tab:parameters}.  
Consequently, we find  that for two models  of  AGN (with and  without
optically thick molecular  torus), Table \ref{tab:parameters} lists at
least lower limits to their luminosities.

In summary, there is no clear explanation for the high opacity seen in
the spectrum of \objectname[]{HE 2347-4342}, especially in regions $C$
and  $D$.   Tests based  on  the  width  of  the \ion{H}{1}  lines are
inconclusive mainly because  of the small number  of  lines available. 
Within the delayed reionization model, we can provide some constraints
on the lifetime, compatible with  currently admitted values, or lower
limits to   the luminosity  of   the AGN causing   the opacity   gaps. 
However, the  large region of  low \ion{H}{1}  opacity  is more easily
explained   in the soft sources   model.    Consequently, 
a search  for $z \sim 2.8$ Lyman  Break  galaxies in this
field would be interesting in order to test the soft sources model,
while a search for AGN with \textit{Chandra} or \textit{XMM-Newton} 
could indicate the presence of a molecular torus.

\subsection{Alternative explanations?}
\label{sec:alternative}

In the  previous sections, we only  considered photoionization  as the
main  process  responsible for  the   different features seen  in  the
spectra of  \objectname[]{HE 2347-4342}.   In \citet{Heap00},  we also
examined two alternative explanations  to the opacity gaps, as arising
in low-density regions or caused by shock-heated gas,  and we will not
consider them any further.   We only note  that the 5 \ion{H}{1} lines
that  appear within the opacity  gaps  have a $\bar{b}_\mathrm{gaps} =
30\pm8~\mbox{km}~\mbox{s}^{-1}$, which  is consistent  with the values
derived in other parts of the spectrum (cf. above). In other words, the
temperature within the opacity gaps  does not seem to be significantly
different  from regions of high  opacity. 
Instead, we can wonder if photoionization is  the only explanation for
the high  opacity seen in  the spectral ranges $C$  and $D$  where few
\ion{H}{1} lines are seen.  Clearly,  such regions are not affected by
low  gas density, since  it  would be difficult  to explain  the large
\ion{He}{2} opacity.    Can collisions ionize  the  \ion{H}{1} without
ionizing \ion{He}{2}?  If this was the case,  we would expect that the
gas be warmer   in regions where   collisional ionization is  the most
important  process. Therefore, we would  expect  that the Doppler  $b$
parameter of the \ion{H}{1} lines in these regions to be significantly
larger than elsewhere.   
In the  previous section, we already  compared
the $b$ parameter of the lines  in regions $C$  and $D$ to the ones in
other regions and found no  significant difference in the median value
of the two  samples.   Within the limits  of  validity of this   test,
mainly due to the small sample of lines, we conclude  that there is no
indication  that collisional  ionization is responsible  for the small
\ion{H}{1} line number density in regions $C$ and $D$.

Finally, it could also be that part of the  observed opacity in region
$C$  and  $D$ is actually    due partially to metal  absorption  lines
associated  either with the   Milky Way or  with intervening  systems. 
However, existing spectra at   longer  wavelengths do not  reveal  any
strong systems at the appropriate  redshifts. Similarly, we do not
expect any strong Galactic lines at $\lambda \sim 1170$\AA.

\section{Conclusions}
\label{sec:summary}

\subsection{Observational results}
\label{sec:summary_observation}

In this paper,  we have   presented    \textit{HST} STIS spectra  of
\objectname[]{HE 2347-4342}. The brightness  of the quasar permits the
use of  the  G140M grating which  provides a   much improved resolution
(0.16  \AA)  compared to the data   obtained by \citet{Reimers97} (0.7
\AA) with the GHRS.  The 2 dimensional nature  of the FUV MAMA detector
allowed us to  better quantify  the   background due  to the sky   and
(mainly)   the  dark    current.   We    confirm   the  findings    by
\citet{Reimers97} that the  spectrum presents regions of high opacity,
some of them devoid of corresponding  \ion{H}{1} lines, and regions of
low opacities (opacity gaps) that match regions displaying a low
number density of \ion{H}{1} lines.

We improve significantly the lower limits on the Gunn-Peterson opacity
in   the  high   opacity    regions.     In  particular,    we    find
$\tau_\mathrm{HeII} =  4.80_{-0.80}^{+\infty}$ in region $C$.  Even at
the lowest  redshift recorded  by the STIS   spectrum, the  opacity is
still $\tau_\mathrm{HeII} =  2.57_{-0.41}^{+0.72}$.  This value   is
significantly larger than the  value of 1.9 measured by \citet{Heap00}
in the spectral range $S$ of the spectrum of \objectname[]{Q 0302-003}
with  a   redshift range    comparable   to   the  STIS  spectrum   of
\objectname[]{HE 2347-4342}.

Another interesting  feature is  the  absence of  the proximity effect
already noticed  by \citet{Reimers97}.   It  is easy to show  that the
large \ion{He}{2}  column density  required to  account for the  large
opacity in the spectral region affected  by the $z_\mathrm{abs} \simeq
z_\mathrm{em}$ absorption also  implies that the amount of \ion{He}{2}
ionizing photons escaping is reduced to a  very small quantity.  It is
interesting  to speculate on   the effects of such   systems on the UV
background radiation.  Of  the 4  objects  for  which the  \ion{He}{2}
Gunn-Peterson absorption has   been  observed,   only two show     the
proximity        effect       (\objectname[]{Q~0302-003}           and
\objectname[]{PKS~1935-692}).           The        other           two
(\objectname[]{HS~1700+6416} and    \objectname[]{HE~2347-4342})  also
show  strong  $z_\mathrm{abs}  \simeq   z_\mathrm{em}$   systems, with
evidence of line   locking \citep[cf.][for HS~1700+6416  and Smette \&
Songaila,  in preparation,  for HE~2347-4342]{TLS97}, which  indicates
that the absorption takes place close to the quasar continuum emitting
region.    If   these 4 quasars    are  representative  of  the quasar
population  as   a whole   --   which is   probably not  the  case, as
\citet{FWP+86} and  \citet{MJ87}) find a dependence  of  the number of
$z_\mathrm{abs} \simeq z_\mathrm{em}$ systems in quasar samples either
on  their radio-loudness   or both  their  radio-loudness  and optical
luminosity --  then one can expect that  only  half of the \ion{He}{2}
ionizing photons escape from  the known quasar population  relative to
the  case  where  all  quasars emit    \ion{He}{2}  ionizing photons.  
Consequently, we  would expect   that the \ion{He}{2}  photoionization
rate  be reduced  by  a  significant factor   in models where   the UV
background radiation  is dominated by quasars,  leading naturally to a
softer UV background.  Another  expected consequence of the  fact that
only a   fraction of the QSOs  actually  provide  \ion{He}{2} ionizing
photons  is that the  UV  background \ion{He}{2} photoionization  rate
should show stronger  fluctuations  than predicted by  current  models
\citep[e.g.][]{Fardal98}, since even fewer quasars would contribute to
the \ion{He}{2} ionizing flux at a given point.

\subsection{Modeling}
\label{sec:summary_modeling}

Comparison between the  \ion{H}{1} and the \ion{He}{2} spectra allowed
us to build a one-dimensional map of the softness of the UV background
radiation.  Except in the opacity  gaps, the observed UV background is
significantly softer  than model  predictions assuming a QSO-dominated
UV   background    \citep[e.g.][]{HM96,Fardal98,Madau99}.  In spectral
ranges $C$ and $D$, it is even larger than  the one predicted for a UV
background whose intensity at 1 Rd has  equal contributions from stars
and AGN \citep[cf.][for  some examples]{Fardal98}. Large values of the
softness parameters   are    in agreement with  the     large observed
\ion{Si}{4}  to  \ion{C}{4}   ratios seen  in  quasar absorption  line
systems.

We have  presented a simple model to  describe the opacity gaps caused
by photoionization due to `hard' sources located  close to the line of
sight to the background quasar. Away from individual absorption lines,
it indicates that  the extent  of the  gap is  related to the   source
luminosity, while, for a given extent, the  amount of transmitted flux
at the center of  the gap is directly related  to the distance of  the
source to the line  of sight.  Finally,  we have expanded the model of
the  \ion{He}{2} Gunn-Peterson  trough  presented in \citet{Heap00} to
account for multiple sources close to the  line of sight. It allows us
to  make   relatively precise   predictions  concerning  the intrinsic
luminosity of  the sources causing  the opacity  gaps, which should be
easily  detectable with  \textit{Chandra} or  \textit{XMM-Newton}. 

We discussed  the possible origin of  the regions  of high \ion{He}{2}
opacity but  low \ion{H}{1} opacity.  We  introduce a new  model where
they are caused by nearby soft  sources.  If only  a small number of
these sources are present close  to the line of  sight, they should be
bright  and  easily observable.   However, we  have few constraints on
their number so that  they could be   much fainter and  more numerous,
with  a space density  comparable with the  Lyman  Break galaxies.  We
compare this model with  the delayed re-ioniation  scenario proposed
by  \citet{Reimers97}.  There is no  clear argument  which leads us to
favor  of one or  the  other model to   explain  the high  \ion{He}{2}
opacity in these regions.  Within  the delayed re-ionization model, if
opacity gaps are indeed caused by AGN, the presence  of a high opacity
region close to  an opacity gap provide  reasonable constraints on the
lifetime or luminosities of the AGN.  On the other hand, soft sources
would more easily explain the large region of low \ion{H}{1} opacity
seen at the same redshift as region $C$.  Optical search for Lyman break
galaxies and X-ray observations can probably settle the issue.

While  this  paper   was being    revised, \citet{KSO+01}    presented
\textit{FUSE}     observations  of   the  line    of    sight  towards
\objectname[]{HE~2347-4342},  covering  a  much longer redshift  range
($2.3 <  z < 2.7$) and twice  higher resolution ($R \sim 15,000$) than
our  \textit{HST} STIS     observations.  However,  over   the  common
wavelength range,   our data show  a larger  signal-to-noise ratio and
especially better controlled systematic   errors  (mainly due to   the
better background determination made possible by the the STIS FUV-MAMA
detector compared to the \textit{FUSE} ones), that allow us to provide
better contraints on the \ion{He}{2} Gunn-Peterson optical depth.  The
behavior  seen in the \textit{HST}   GHRS and STIS observations, i.e.,
alternance  of regions of  low and high opacity  is confirmed and seen
over  the whole     spectral  range covered   by     the \textit{FUSE}
observations.  In  the   \textit{FUSE} spectrum, the   optical  depths
measured  over spectral  ranges of  $\sim  5$\AA\,  appear to converge
towards theoretically predicted values \citep[e.g.][]{Fardal98,mbm+00}
at $z  < 2.75$ but are systematically  larger than predictions  in the
redshift range  common  to both \textit{FUSE}  and  STIS observations. 
Comparison    between  the   \textit{FUSE}     observations   and  the
corresponding Keck--HIRES \ion{H}{1}  spectrum shows that $\eta \equiv
N_\mathrm{HeII}/N_\mathrm{HI}$  varies  between $\sim  1$  to at least
several hundreds,  as in the  STIS spectral range.  In accord with the
findings of this paper, \citet{KSO+01} 
conclude   that  these regions   of large  $\eta$   (or
softness  parameter  $S \propto \eta$)  are  subject to  a much softer
background  radiation    either  because    of    delayed  \ion{He}{2}
re-ionization  or due to the presence  of `soft' sources  close to the
line-of-sight. Although  both  explanations are equally  possible over
some spectral range, the fact  that excursions of $\eta$ towards large
values continue to $ z \sim 2.3$ probably favour the latter one.



\acknowledgments 

We thank  Joop Schaye,  Mark Fardal  and Gerry Kriss  for discussions. 
Critical comments from an anonymous  referee are also acknowledged. We
thank Michel Fioc for pointing out that  the incomplete Gamma function
can  be used in   Eq.    \ref{eq:l_cloud_analytical}.  This work   was
supported by NASA Guaranteed Time Observer funding to the STIS Science
Team.


\clearpage

\clearpage





\clearpage


\begin{deluxetable}{ccccc}
\tablecolumns{5}
\tablecaption{Measured \ion{He}{2} opacities
  \label{tab:opacity}
  }
\tablehead{
  \colhead{Spectral range}       &
  \colhead{Wavelength}           & 
  \colhead{Mean}                 & 
  \colhead{Mean}                 &
  \colhead{$\tau_\mathrm{HeII}$\tablenotemark{a}} \\
  \colhead{}               &
  \colhead{range}          & 
  \colhead{redshift}       & 
  \colhead{normalized}     &
  \colhead{} \\
  \colhead{}               &
  \colhead{(\AA)}          & 
  \colhead{}               & 
  \colhead{flux}      &
  \colhead{} \\
}
\startdata
  A & 1175.60--1178.00 & 2.8738 & $ \phantom{-}0.018\pm0.009 $ & $ 4.00^{+0.67}_{-0.40}$\\
  B & 1174.00--1174.90 & 2.8661 & $ \phantom{-}0.930\pm0.040 $ & $ 0.07^{+0.04}_{-0.04}$\\
  C & 1170.00--1173.50 & 2.8572 & $ \phantom{-}0.008\pm0.010 $ & $ 4.80^{+\infty}_{-0.80}$ \\
  D & 1165.00--1169.95 & 2.8431 & $           -0.009\pm0.011 $ & $> 4.54$ \\
  E & 1162.00--1165.00 & 2.8300 & $ \phantom{-}0.057\pm0.018 $ & $ 2.86^{+0.39}_{-0.28}$\\
  F & 1159.00--1161.00 & 2.8185 & $ \phantom{-}0.642\pm0.038 $ & $ 0.44^{+0.06}_{-0.06}$\\
  G & 1157.75--1158.25 & 2.8119 & $ \phantom{-}0.783\pm0.090 $ & $ 0.24^{+0.12}_{-0.11}$\\
  H & 1155.00--1157.75 & 2.8066 & $ \phantom{-}0.096\pm0.032 $ & $ 2.34^{+0.40}_{-0.28}$\\
  I & 1150.00--1154.95 & 2.7938 & $ \phantom{-}0.024\pm0.030 $ & $ 3.72^{+\infty}_{-0.81}$ \\
  J & 1145.00--1149.95 & 2.7773 & $ \phantom{-}0.076\pm0.039 $ & $ 2.57^{+0.72}_{-0.41}$\\
\hline\\
C+D & 1165.00--1173.50 & 2.8490 & $           -0.001\pm0.007 $ & $> 4.90$ \\
I+J & 1145.00--1154.95 & 2.7856 & $ \phantom{-}0.049\pm0.025 $ & $ 3.02^{+0.70}_{-0.41}$\\
\enddata 
\tablenotetext{a}{
  If the mean intensity $\bar{I}$ is negative, the quoted lower limit
  to the opacity is  $- \log{\sigma_\mathrm{I}}$.
}
\end{deluxetable}         
\begin{deluxetable}{cccccccl}
  \tablecolumns{8}
  \tablecaption{Model parameters
    \label{tab:parameters}
    }
  \tablehead{
    \colhead{Source}                       &
    \colhead{$z_s$}                          &
    \colhead{$L_\mathrm{HI}^s$\tablenotemark{a}}             &
    \colhead{$L_\mathrm{HeII}^s$\tablenotemark{a}}             &
    \colhead{$f_\mathrm{HI}^s$\tablenotemark{a}}             &
    \colhead{$f_\mathrm{HeII}^s$\tablenotemark{a}}            &
    \colhead{$p^{s}$}           &
    \colhead{Comments}                      \\
    }
  \startdata
    &   2.885    & $220$  & $ .  $   & 250 &   .   &   0   & HE 2347-4342\\
  1 &   2.81500  & $4.1$  & $0.25$   & 5.0 &  5.0  &   4.2 &  hard source (AGN)\\
  2 &   2.86609  & $1.3$  & $0.08$   & 1.5 &  1.5  &   1.0 &  hard source (AGN)\\
  3 &   2.79583  & $0.4$  &   .      & 0.5 &  .    &   0.8 &  soft source  \\
  4 &   2.83000  & $0.4$  &   .      & 0.5 &  .    &   0.4 &  soft source  \\
  5 &   2.83200  & $0.4$  &   .      & 0.5 &  .    &   0.4 &  soft source  \\
  6 &   2.84100  & $0.4$  &   .      & 0.5 &  .    &   0.4 &  soft source  \\
  7 &   2.84300  & $0.4$  &   .      & 0.5 &  .    &   0.4 &  soft source  \\
  8 &   2.85000  & $0.4$  &   .      & 0.5 &  .    &   0.4 &  soft source  \\
  9 &   2.85800  & $0.8$  &   .      & 1.0 &  .    &   0.4 &  soft source  \\
  \enddata
  \tablenotetext{a}{
    Fluxes              assumed           $H_\mathrm{0}              =
    65~\mbox{km}~\mbox{s}^{-1}~\mbox{Mpc}^{-1}$, $\Omega_\mathrm{M}  =
    0.3$ and $\Lambda =  0.7$. The  sources  are assumed to present  a
    flat spectrum in $f_\lambda$. Luminosities  $L$ are given in units
    of      $10^{30}~\mbox{ergs}~\mbox{s}^{-1}~\mbox{Hz}^{-1}$, fluxes
    $f_\lambda$          are       given       in       units       of
    $10^{-17}~\mbox{ergs}~\mbox{s}^{-1}~\mbox{\AA}^{-1}$,  and  impact
    parameters  $p$ in   comoving  Mpc.   The  parameters  are    well
    constrained only for the  `hard'  sources which cause the  opacity
    gaps. Values at \ion{He}{2} Lyman limit for   \objectname[]{HE
      2347-4342}  are nil due to the intrinsinc absorbers. 
    }
\end{deluxetable}
%
%
%
%


\end{document}